\documentclass[12pt]{iopart}
\usepackage{iopams}
\usepackage{graphicx}
\usepackage{color}

\begin{document}

\title{Liquid crystal director fields in three-dimensional non-Euclidean geometries}

\author{Jean-Fran\c cois Sadoc$^1$, R\'emy Mosseri$^2$, and Jonathan~V.~Selinger$^3$}

\address{$^1$Laboratoire de Physique des Solides (CNRS-UMR 8502), B{\^a}t. 510, Universit{\'e} Paris-sud (Paris-Saclay), F 91405 Orsay cedex, France}
\address{$^2$Laboratoire de Physique Th\'eorique de la Mati\`ere Condens\'ee, Sorbonne Universit\'e, CNRS UMR 7600, F-75005 Paris, France}
\address{$^3$Department of Physics, Advanced Materials and Liquid Crystal Institute, Kent State University, Kent, Ohio 44242, USA}
\eads{\mailto{jean-francois.sadoc@u-psud.fr}, \mailto{remy.mosseri@upmc.fr}, \mailto{jselinge@kent.edu}}
\vspace{10pt}
\begin{indented}
\item[]\today
\end{indented}

\begin{abstract}
This paper investigates nematic liquid crystals in three-dimensional curved space, and determines which director deformation modes are compatible with each possible type of non-Euclidean geometry.  Previous work by Sethna \emph{et al.}\ showed that double twist is frustrated in flat space $\mathbb{R}^3$, but can fit perfectly in the hypersphere $\mathbb{S}^3$.  Here, we extend that work to all four deformation modes (splay, twist, bend, and biaxial splay) and all eight Thurston geometries.  Each pure mode of director deformation can fill space perfectly, for at least one type of geometry.  This analysis shows the ideal structure of each deformation mode in curved space, which is frustrated by the requirements of flat space.
\end{abstract}

\section{Introduction}

In a liquid crystal, the ground state does not necessarily have a uniform director field.  Rather, the ground state of the local free energy may have some optimal gradient of the director field.  For example, if the liquid crystal is chiral, the free energy favors a twist deformation of the director.  If the liquid crystal is composed of bent-core molecules, the free energy may favor a bend deformation~\cite{Dozov2001,Jakli2018}.

Whenever the optimal director field is not uniform, one must ask:  Is it possible to have the optimal director deformation everywhere in space?  In many cases, the answer is no.  In blue phases, the optimal double twist cannot occur everywhere in space, and hence the liquid crystal forms a lattice of double-twist tubes separated by disclination lines.  Likewise, in bent-core liquid crystals, the optimal bend cannot occur by itself everywhere in space, and hence the liquid crystal forms a heliconical twist-bend nematic ($N_{TB}$) phase.  Both of these cases can be regarded as examples of geometric frustration, in which a system develops a complex global structure because the optimal local structure cannot fill up space~\cite{Sadoc1999,Kamien2001,Nelson2002b}.

Many years ago, Sethna \emph{et al.}~\cite{Sethna1983} suggested an interesting theoretical approach to analyze geometric frustration in blue phases.  They pointed out that the ideal double twist can fit perfectly in the three-dimensional (3D) curved non-Euclidean geometry of a hypersphere, represented mathematically as $\mathbb{S}^3$, with the appropriate curvature radius.  The structure experiences frustration when it is forced to exist in flat Euclidean space, denoted as $\mathbb{R}^3$.  Hence, the network of disclination lines in a blue phase can be understood as a result of projecting the structure from $\mathbb{S}^3$ into $\mathbb{R}^3$.  In that way, the structure of blue phases is analogous to the packing of spherical particles~\cite{Sadoc1999,Nelson2002b}, which has an ideal icosahedral order that can fit perfectly in $\mathbb{S}^3$ but is frustrated in $\mathbb{R}^3$.

More recently, Niv and Efrati~\cite{Niv2018} developed a systematic method to analyze geometric compatibility in 2D liquid crystals.  Their method shows that an optimal splay $S$ and optimal bend $B$ can fit perfectly in a 2D surface if $S^2+B^2=-K_G$, where $K_G$ is the Gaussian curvature of the surface.  Hence, a nonzero splay or bend can exist everywhere in a 2D non-Euclidean geometry with the appropriate negative curvature, but must be frustrated in a flat Euclidean plane and on a sphere.  This analysis has not yet been extended from 2D to 3D.

The purpose of this article is to generalize the theory of Sethna \emph{et al.}\ to the other director deformation modes, and determine whether each mode can fit perfectly in some 3D curved space.  To classify the director deformation modes, we use the mathematical analysis of Machon and Alexander~\cite{Machon2016}, further discussed by Selinger~\cite{Selinger2018}.  This work shows that there are four distinct modes:  splay, twist, bend, and a fourth mode that might be called ``biaxial splay.''  (In this approach, pure twist means double twist, while cholesteric single twist is a combination of twist and biaxial splay~\cite{Selinger2018}.)  This analysis has already been used to characterize the director deformations that can fill up flat Euclidean space~\cite{Virga2019}.  To classify the possible types of curved space, we use the analysis of Thurston~\cite{Thurston1997}, which shows that there are eight possible homogeneous geometries in 3D.  For each of these geometries, we construct simple director fields, calculate the deformation modes, and identify cases where a single pure mode can have a constant nonzero value.

The main result of this study is that each of the pure modes can exist in some non-Euclidean geometry.  Pure twist can exist in the hypersphere $\mathbb{S}^3$, as shown by Sethna \emph{et al.}, and also in the spaces $\widetilde{SL}(2,R)$ and $Nil$.  Similarly, pure splay can exist in the hyperbolic geometry $\mathbb{H}^3$, and pure bend in $\mathbb{H}^3$, $\mathbb{H}^2 \times \mathbb{R}$, and $Sol$.  Pure biaxial splay can exist in the spaces $\widetilde{SL}(2,R)$, $Nil$, and $Sol$.  These results have implications for geometric frustration of liquid-crystal phases.  For example, if a liquid crystal is composed of bent-core molecules, it can have an optimal director field with pure constant bend.   Likewise, if a liquid crystal is composed of pear-shaped molecules, it can have an optimal director field with pure constant splay.  Either of these structures fits perfectly in $\mathbb{H}^3$, but is frustrated when forced to exist in flat space $\mathbb{R}^3$.  This frustration leads to the formation of modulated structures with defect lines, or to mixtures of the optimal mode with other deformation modes.  Understanding the ideal non-Euclidean structure, even as a mathematical abstraction, provides insight into these more complex structures that can form experimentally.

The plan of this paper is as follows. In Sec. 2, we set up the formalism for director deformation modes and the Oseen-Frank free energy in non-Euclidean geometry.  In Sec. 3, we discuss examples in 2D curved surfaces. In Sec. 4, we go on to 3D curved space, consider each of the eight Thurston geometries, and determine what pure director deformations are permitted.

\section{Formalism for nematic order in curved space}

The theory of liquid-crystal order in 2D non-Euclidean geometry has been developed by many investigators over thirty years, and some of the history can be briefly summarized as follows.  Nelson and Peliti~\cite{Nelson1987} introduced the concept of hexatic order in fluctuating membranes, and derived a necessary coupling of membrane shape with orientational order and defects.  Park \emph{et al.}~\cite{Park1992} extended the concept to general $n$-atic order on closed vesicles, which must have $2n$ defects of topological charge $1/n$, which distort the vesicle shape away from a sphere.  Nelson~\cite{Nelson2002} pointed out that nematic order on a sphere could be exploited to create a tetravalent colloidal chemistry, which might be useful for photonic applications.  Fernandez-Nieves \emph{et al.}~\cite{Fernandez-Nieves2007} developed experimental realizations of spherical liquid crystals by fabricating nematic shells, with water both inside and outside.  They found defect structures more complex than expected, because the inner droplet was consistently off-center.  Shin \emph{et al.}~\cite{Shin2008} explored the effects of unequal Frank elastic constants for splay and bend, and found that a difference of elastic constants can shift the positions of the defects.  Nguyen \emph{et al.}~\cite{Nguyen2013} investigated the extrinsic coupling of orientational order and curvature, which arises because a curved 2D membrane is embedded in 3D space, and found that this extrinsic coupling greatly changes the predictions for vesicle shapes.  Keber \emph{et al.}~\cite{Keber2014} extended the theory from equilibrium to active nematic vesicles, and showed that active defects move in complex trajectories around a sphere.

In this section, we generalize the theoretical formalism from a 2D curved surface to a 3D curved space.  The approach is mostly the same as in the previous work on 2D curved surfaces, except that we use a different basis for the director field, as discussed below.

\subsection{Differential geometry}

A 3D curved space is described by three coordinates, which are conventionally written as $\sigma^1$, $\sigma^2$, and $\sigma^3$.  Any position in the space can be written as $\mathbf{R}(\sigma^1,\sigma^2,\sigma^3)$, and tangent vectors along the field lines are $\mathbf{t}_i=\partial_i \mathbf{R}$.  These vectors form a basis for the space, and they are called the covariant basis vectors.  Distances between nearby points are given by $(ds)^2 = g_{ij} d\sigma^i d\sigma^j$, where $g_{ij}=\mathbf{t}_i\cdot\mathbf{t}_j$ is the covariant metric tensor.  Its inverse is the contravariant metric tensor $g^{ij}$, with $g^{ij}g_{jk}=\delta^i_k$.  Using $g^{ij}$, we can define the contravariant basis vectors $\mathbf{t}^i=g^{ij}\mathbf{t}_j$.

We must consider how the covariant and contravariant basis vectors vary as functions of position.  Here, we assume that the curved space is not embedded in any higher-dimensional space, so there is no extrinsic variation outside of the space.  The intrinsic variation within the space is $\partial_i\mathbf{t}_j={\Gamma^k}_{ij}\mathbf{t}_k$ and $\partial_i\mathbf{t}^k=-{\Gamma^k}_{ij}\mathbf{t}^j$, with the Christoffel symbols ${\Gamma^k}_{ij}$ given by
\begin{equation}
{\Gamma^k}_{ij}=\mathbf{t}^k\cdot\partial_i\mathbf{t}_j
=\frac{1}{2}g^{kl}\left(\partial_i g_{jl}+\partial_j g_{il}-\partial_l g_{ij}\right).
\label{equa1}
\end{equation}

We apply this formalism to homogeneous 3D spaces.  Homogeneous means that all points are equivalent under a global isometry. However, homogeneity does not mean isotropy. Although all points are equivalent, the curvature properties may differ when considering different directions emerging from the points. If the curvature is the same for all directions, the space is said to be isotropic; otherwise it is anisotropic and homogeneous.

The intrinsic curvature of a manifold $M$ is characterized by the Riemann curvature tensor
\begin{equation}
{R^a}_{bcd}=\partial_c {\Gamma^a}_{bd} - \partial_d {\Gamma^a}_{bc} + {\Gamma^a}_{ce}{\Gamma^e}_{bd} - {\Gamma^a}_{de}{\Gamma^e}_{bc}.
\end{equation} 
A simpler description is given by the Ricci tensor $R_{ij}$, which is a contracted form of the Riemann tensor, with $R_{ij}={R^k}_{ikj}$. In 3D, the Ricci tensor still encompasses full information about the curvature; this is not true in higher dimensions. A further contraction leads to the coarser ``scalar" curvature $R=g^{ij}R_{ij}$.

Another important quantity is the ``sectional curvature" $K_S(\sigma)$, where $\sigma$ denotes a 2D plane in the 3D tangent space of the manifold. Briefly, $K_S(\sigma)$ measures the Gaussian curvature of a 2D submanifold of $M$ whose tangent space coincides with $\sigma$. (More precisely, this 2D submanifold is the ruled surface in $M$, union of geodesics tangent to this plane.) Now the scalar $R$ corresponds to an average of $K_G$ over all possible planes in the tangent space. Isotropic spaces of constant scalar curvature show equal sectional curvature for any plane $\sigma$.

A way to characterize the anisotropy is to consider the eigenvalues of the mixed tensor ${R^i}_j=g^{ik}R_{kj}$, whose trace $R^i_i$ gives the scalar curvature. Each eigenvalue relates to the average sectional curvature, in the corresponding direction, for all planes sharing that direction. Inequality of the eigenvalues is a signature of anisotropy. The list of eigenvalues will be denoted here as $diag({R^i}_j)$, and the corresponding eigenvectors as $\{\hat{\mathbf{a}},\hat{\mathbf{b}},\hat{\mathbf{c}}\}$.

\subsection{Director field}

In a nematic phase, the molecules have orientational order along an axis, called the director, given by the unit vector $\hat{\mathbf{n}}$.  Both $+\hat{\mathbf{n}}$ and $-\hat{\mathbf{n}}$ represent the same physical state, with order along the same axis.  In general, the axis of orientational order varies smoothly as a function of position, and hence it is written as the director field $\hat{\mathbf{n}}(\sigma^1,\sigma^2,\sigma^3)$.  The Oseen-Frank free energy gives the elastic free energy cost associated with spatial variations of the director field.  To calculate the free energy, we must take derivatives of $\hat{\mathbf{n}}$.  To find these derivatives, we must express $\hat{\mathbf{n}}$ in terms of components along some local basis.  Of course, the results for the free energy must be independent of the choice of basis.

In the literature on liquid crystals on 2D curved surfaces, researchers generally construct a local \emph{orthonormal} basis at every point on the surface, and express $\hat{\mathbf{n}}$ in terms of the two basis vectors $\hat{\mathbf{v}}_1$ and $\hat{\mathbf{v}}_2$.  This choice of basis is particularly convenient in 2D, because one can write $\hat{\mathbf{n}}=\hat{\mathbf{v}}_1\cos\theta+\hat{\mathbf{v}}_2\sin\theta$, where $\theta$ is a local angle representing the direction of orientational order.  Gradients of $\hat{\mathbf{n}}$ can then be expressed in terms of gradients of $\theta$ and gradients of the basis vectors, given by the connection $A_i=\hat{\mathbf{v}}_1\cdot\partial_i\hat{\mathbf{v}}_2=-\hat{\mathbf{v}}_2\cdot\partial_i\hat{\mathbf{v}}_1$.  Nelson and Peliti~\cite{Nelson1987} showed that the connection $A_i$ couples with gradients of the angle field $\theta$, just as the magnetic vector potential couples with gradients of the phase field in a superconductor.

When extending the theory from 2D to 3D, we use two other choices of basis.  First and most generally, we use the conventional covariant or contravariant basis of tangent vectors.  With the covariant basis vectors $\mathbf{t}_i=\partial_i \mathbf{R}$, we can write $\hat{\mathbf{n}}=n^i \mathbf{t}_i$, where $n^i$ are the contravariant components of the director.  Equivalently, with the contravariant basis vectors $\mathbf{t}^i=g^{ik}\mathbf{t}_k$, we can also write $\hat{\mathbf{n}}=n_i \mathbf{t}^i$, and $n_i=g_{ij}n^j$ are the covariant components of the director.  In this notation, it is essential to remember that the basis vectors are generally \emph{not} orthonormal, and hence the director components are not sines and cosines.  Rather, the director components are normalized as $n^i n_i=g_{ij}n^i n^j=g^{ij}n_i n_j=1$.  Gradients of $\hat{\mathbf{n}}$ can then be written as
\begin{eqnarray}
\partial_i\hat{\mathbf{n}}&=&\partial_i(n^j \mathbf{t}_j) = \mathbf{t}_k (\partial_i n^k+{\Gamma^k}_{ij}n^j)=\mathbf{t}_k D_i n^k\nonumber\\
&=&\partial_i(n_k \mathbf{t}^k)=\mathbf{t}^j (\partial_i n_j-{\Gamma^k}_{ij}n_k)=\mathbf{t}^j D_i n_j.
\label{equa2}
\end{eqnarray}
These equations define the covariant derivatives $D_i n_j$ and $D_i n^k$.

Second, if the curvature of the space is anisotropic, it is useful to distinguish among the special directions associated with the curvature.  For that purpose, we calculate the orthonormal eigenvectors of the mixed Ricci tensor, and use that set of vectors $\{\hat{\mathbf{a}},\hat{\mathbf{b}},\hat{\mathbf{c}}\}$ as our basis.  The director field can then be characterized by an azimuthal angle $\alpha$ and polar angle $\beta$ in the $\{\hat{\mathbf{a}},\hat{\mathbf{b}},\hat{\mathbf{c}}\}$ frame,
\begin{equation} 
\label{eqn:field}
\hat{\mathbf{n}}(\alpha,\beta)=  \sin \beta \cos \alpha \,\hat{\mathbf{a}} +\sin \beta \sin \alpha\,\hat{\mathbf{b}}+ \cos \beta\,\hat{\mathbf{c}}. 
\end{equation}
Whenever the mixed Ricci tensor is uniaxial, with two degenerate eigenvalues that are distinct from the third, we let $\hat{\mathbf{c}}$ be the eigenvector corresponding to the non-degenerate eigenvalue.  In the following sections, we will often construct a director field using this basis, and then find the director components and calculate the director gradient tensor $D_i n_j$ in the conventional basis.

As derived by Machon and Alexander~\cite{Machon2016}, and discussed by Selinger~\cite{Selinger2018}, the director gradient tensor can be decomposed into four parts, which represent physically distinct types of spatial variations.  In covariant notation, this decomposition becomes
\begin{equation}
D_i n_j = -n_i B_j + \frac{1}{2}T\epsilon_{ijk}n^k + \frac{1}{2}S(g_{ij}-n_i n_j) + \Delta_{ij} .
\label{equa3}
\end{equation}
The first term is \emph{bend} of the director field, given by the vector
\begin{equation}
B_j=-n^i D_i n_j .
\label{equa4}
\end{equation}
Because $\hat{\mathbf{n}}$ is a unit vector, the bend vector is perpendicular to $\hat{\mathbf{n}}$,
with $n^j B_j = 0$.  If $\hat{\mathbf{n}}$ is tangent to geodesic lines, then its covariant derivative vanishes and hence its bend vanishes.  The second term is \emph{twist} (i.e.\ double twist), represented by the pseudoscalar
\begin{equation}
T=\epsilon^{lij}n_l D_i n_j .
\label{equa5}
\end{equation}
In the twist term, the covariant Levi-Civita tensor $\epsilon_{ijk}$ includes a factor of $\sqrt{|g|}$, and the contravariant Levi-Civita tensor $\epsilon^{lij}$ includes a factor of $1/\sqrt{|g|}$, with $g=\det g_{ij}$.  The third term is \emph{splay} (i.e.\ double splay), defined by the scalar
\begin{equation}
S=g^{ij}D_i n_j=D_j n^j.
\label{equa6}
\end{equation}
The fourth term has been called \emph{biaxial splay}~\cite{Selinger2018}, and can be expressed as the tensor
\begin{equation}
\Delta_{ij}=\frac{1}{2}\left[D_i n_j+D_j n_i+n_i B_j+n_j B_i-S(g_{ij}-n_i n_j)\right].
\label{equa7}
\end{equation}
It is a symmetric, traceless tensor in the plane perpendicular to $\hat{\mathbf{n}}$, so that $\Delta_{ij}=\Delta_{ji}$, $g^{ij}\Delta_{ij}=0$, and $n^i\Delta_{ij}=0$.  The eigenvalues of the tensor ${\Delta^i}_j = g^{ik}\Delta_{kj}$ are $0$, positive, and negative.  The eigenvalue $0$ corresponds to the eigenvector $\hat{\mathbf{n}}$, and the other two eigenvalues correspond to two vectors perpendicular to $\hat{\mathbf{n}}$, which show the directions in which the director field splays outward and inward.  This combination of outward and inward gradients is characteristic of biaxial splay.

\subsection{Free energy}

In covariant notation, the simplest possible version of the Oseen-Frank elastic free energy density of a nematic liquid crystal is
\begin{eqnarray}
F&=&\frac{1}{2}K g^{ik}\partial_i\hat{\mathbf{n}}\cdot\partial_k\hat{\mathbf{n}}=\frac{1}{2}K g^{ik}g^{jl}(D_i n_j)(D_k n_l)\nonumber\\
&=&\frac{1}{4}K S^2 +\frac{1}{4}K T^2 +\frac{1}{2}K B^j B_j +\frac{1}{2}K\Delta^{ij}\Delta_{ij}.
\label{equa8}
\end{eqnarray}
In this expression, the four modes of splay, twist, bend, and biaxial splay contribute equally to the elastic free energy (except for factors of $\frac{1}{2}$ and $\frac{1}{4}$).  For that reason, this expression is called the single-elastic-constant approximation.  More generally, the four modes may have four distinct elastic coefficients, as discussed in the review article~\cite{Selinger2018}.  In the following sections, we will describe some interesting director fields, and characterize their bend, splay, twist and biaxial splay content by the scalars $S^2$, $T^2$, $|\mathbf{B}|^2=B^j B_j$, and
$\Tr(\Delta^2)=\Delta^{ij}\Delta_{ij}$, which enter the elastic free energy.

If a liquid crystal is made of chiral molecules, then it does not have any reflection symmetry.  In that case, the free energy must have an additional term that is linear in the twist, with the form $-KqT$.  By completing the square and subtracting an unimportant constant, we can rewrite the free energy density as
\begin{equation}
F_\mathrm{chiral}=\frac{1}{4}K S^2 +\frac{1}{4}K (T-2q)^2 +\frac{1}{2}K B^j B_j +\frac{1}{2}K\Delta^{ij}\Delta_{ij}.
\label{equa9}
\end{equation}
Clearly the \emph{local} free energy density is minimized when $T=2q$, $S=0$, $B_j=0$, and $\Delta_{ij}=0$.  The question then becomes:  Is it possible to find any \emph{global} director field with those values of $T$, $S$, $B_j$, and $\Delta_{ij}$?  For a liquid crystal in 3D Euclidean space, the answer is no; there is no global director field with constant nonzero double twist, and all other modes equal to zero.  Because of this geometric frustration, a chiral liquid crystal must form a more complex structure.  In most cases, a chiral liquid crystal forms a cholesteric phase, which has constant nonzero $T$ and $\Delta_{ij}$, but zero $S$ and $B_j$.  Alternatively, it may form a blue phase, with a lattice of double twist tubes separated by disclination lines, and the ideal structure at the center of each tube.  The insight of Sethna \emph{et al.}~\cite{Sethna1983} was that the ideal local double twist structure can be achieved everywhere in a specific 3D non-Euclidean geometry, which is a hypersphere of radius $1/q$.  We will discuss this structure in Sec.~4.2 below.

Apart from chirality, liquid crystals can also be made of molecules with other types of asymmetry.  For example, bent-core liquid crystals are composed of banana-shaped molecules, which have a long molecular axis and a transverse orientation.  If the transverse orientation is ordered, then the long molecular axis has a tendency to bend, known as the converse bend flexoelectric effect~\cite{Jakli2018,Meyer1969}.  This tendency can be described by the free energy density
\begin{equation}
F_\mathrm{banana}=\frac{1}{4}K S^2 +\frac{1}{4}K T^2 +\frac{1}{2}K(B^j-\bar{B}^j)(B_j-\bar{B}_j) +\frac{1}{2}K\Delta^{ij}\Delta_{ij},
\label{equa10}
\end{equation}
where $\bar{B}_j$ is the spontaneous bend, which is a vector of constant magnitude, perpendicular to $\hat{\mathbf{n}}$.  This local free energy density is minimized when $B_j=\bar{B}_j$, $S=0$, $T=0$, and $\Delta_{ij}=0$.  This structure experiences geometric frustration in 3D Euclidean space, because no global director field has those derivatives.  One possible resolution is to form a heliconical twist-bend nematic ($N_{TB}$) phase, with constant nonzero $B$, $T$, and $\Delta_{ij}$, but zero $S$.  Following the example of Sethna \emph{et al.}, we will look for a non-Euclidean geometry in which the ideal structure of pure bend can form without frustration.

Two further possibilities can also be considered.  A liquid crystal can be composed of pear-shaped molecules, with a long molecular axis and a polar orientation from the narrow end to the wide end.  If the polar orientation is ordered, then the long molecular axis has a tendency to splay, known as the converse splay flexoelectric effect~\cite{Jakli2018,Meyer1969}.  That tendency can be represented by the free energy density
\begin{equation}
F_\mathrm{pear}=\frac{1}{4}K (S-\bar{S})^2 +\frac{1}{4}K T^2 +\frac{1}{2}K B^j B_j +\frac{1}{2}K\Delta^{ij}\Delta_{ij},
\label{equa11}
\end{equation}
where $\bar{S}$ is the spontaneous splay.  Likewise, a liquid crystal can be composed of more complex biaxial molecules, perhaps with an elongated tetrahedral shape.  That shape could give a tendency toward the biaxial splay mode, expressed by the free energy density
\begin{equation}
F_\mathrm{biax}=\frac{1}{4}K S^2 +\frac{1}{4}K T^2 +\frac{1}{2}K B^j B_j
+\frac{1}{2}K(\Delta^{ij}-\bar{\Delta}^{ij})(\Delta_{ij}-\bar{\Delta}_{ij}).
\label{equa12}
\end{equation}
It has a spontaneous deformation $\bar{\Delta}_{ij}$, which is a symmetric, traceless tensor with constant eigenvalues, in the plane perpendicular to $\hat{\mathbf{n}}$.  Both of these forms for the free energy density also have geometric frustration in 3D Euclidean space.  We will look for non-Euclidean geometries in which their ideal local structures can fill space without frustration.

\section{Director fields in 2D surfaces of constant curvature}

The only possible deformations in 2D are splay and bend.  Niv and Efrati derived the ``compatibility" constraint $S^2+B^2=-K_G$, which shows that a regular field configuration with non-vanishing constant splay or bend can exist in a negatively curved surface, but not in a surface of zero or positive curvature~\cite{Niv2018}.  We illustrate this phenomenon with specific examples, mainly as an introduction to further analysis of the 3D case.

In 2D, curvature is captured by one intrinsic quantity, the Gaussian curvature, and there are three homogeneous surfaces of constant curvature:  the Euclidean plane $\mathbb{R}^2$, the sphere $\mathbb{S}^2$ and the hyperbolic plane $\mathbb{H}^2$. These 2D surfaces are all isotropic. Notice that $\mathbb{S}^2$ has a simple visualization, because it can be isometrically embedded in $\mathbb{R}^3$, while the representation of $\mathbb{H}^2$ is more complex.

There are many ways to define a vector field on a manifold.  We consider a simple class of vector fields, which are defined as follows:  Suppose that a surface is represented as $\mathbf{R}(u,v)$, where $u$ and $v$ are two independent coordinates.  One vector field is defined as $\mathbf{t}_u=\partial_u \mathbf{R}$, normalized to unit magnitude.  It is tangent to field lines $\mathbf{R}(u,v)$ with variable $u$ and constant $v$.  Another vector field is defined as $\mathbf{t}_v=\partial_v \mathbf{R}$, also normalized to unit magnitude.  It is tangent to field lines $\mathbf{R}(u,v)$ with constant $u$ and variable $v$.  These two cases will be interesting enough to demonstrate pure splay or bend.  

As mentioned in the previous section, vector fields tangent to geodesic lines have zero bend.  Hence, we can only find nonzero bend if the vector field is tangent to field lines that are not geodesics.

We briefly discuss the cases of $\mathbb{R}^2$ and $\mathbb{S}^2$, and then concentrate on the most interesting case of $\mathbb{H}^2$.

\subsection{Euclidean plane $\mathbb{R}^2$}

The simplest choice of coordinate system is cartesian, with metric $d\ell^2=dx^2+dy^2$.  In that case, we consider a director field aligned along either the $x$ or $y$ direction.  This director field is constant, with zero splay and zero bend.

Another possible choice is polar coordinates $(\rho,\theta)$, with metric $d\ell^2=d\rho^2+\rho^2 d\theta^2$.  With this coordinate system, we can consider two types of director field.  First, varying $\rho$ with constant $\theta$ leads to radial field lines.  The corresponding director field $n^i=(1,0)$ has nonuniform splay $S = 1/\rho$ and zero bend.  Alternatively, varying $\theta$ with constant $\rho$ leads to circular field lines.  The corresponding director field $n^i=(0,1/\rho)$ has nonuniform bend $B^i = (1/\rho,0)$ with magnitude $|\mathbf{B}|^2=1/\rho^2$ and zero splay.  Note that this coordinate system (and hence the derived director fields) is singular at the origin.

\subsection{Unit sphere $\mathbb{S}^2$}

We consider spherical coordinates $(\theta,\phi)$, with metric $d\ell^2=d\theta^2+\sin^2\theta\,d\phi^2$, from which the standard embedding of $\mathbb{S}^2$ in $\mathbb{R}^3$ is derived.  With this coordinate system, we can construct two types of director field.  First, varying $\theta$ with constant $\phi$ leads to meridian lines.  The corresponding director field $n^i=(1,0)$ has nonuniform splay $S=\cot\theta$ and zero bend (as expected because the meridians are geodesics).  Alternatively, varying $\phi$ with constant $\theta$ leads to parallel circles on the sphere.  The corresponding director field $n^i=(0,\csc\theta)$ has nonuniform bend $B^i = (\cot\theta,0)$ with magnitude $|\mathbf{B}|^2=\cot^2 \theta$ and zero splay.

Like the polar coordinate system for $\mathbb{R}^2$, the spherical coordinate system for $\mathbb{S}^2$ has singularities, which occur at the two poles.  For $\mathbb{S}^2$, the occurrence of singularities is not an artifact of this coordinate system, because of the well-known ``hairy ball theorem,'' which shows that any vector field tangent to $\mathbb{S}^2$ must have at least one singularity.

\subsection{Hyperbolic plane $\mathbb{H}^2$}

The most interesting 2D case is provided by the hyperbolic plane $\mathbb{H}^2$.  As a first step, we must construct a coordinate system for $\mathbb{H}^2$, and visualize the field lines corresponding to these coordinates.  This task is challenging because there is no isometric embedding of the hyperbolic plane $\mathbb{H}^2$ into Euclidean space $\mathbb{R}^3$.  However, it is possible to represent $\mathbb{H}^2$ as an hyperbolic surface of equation $x_1^2+x_2^2-x_3^2=-1$, provided that the 3D embedding space is given a Minkowski metric $d\ell^2=dx_1^2+dx_2^2-dx_3^2$.  Among the different ways to parameterize that embedded surface, we choose the exponential coordinate system $(\sigma,\mu)$ defined by~\cite{Costa2001}
\begin{equation}
x_1 = \mu  e^{-\sigma},\
x_2 = \frac{1}{2} \mu ^2 e^{-\sigma}+\sinh (\sigma),\
x_3 = \frac{1}{2} \mu ^2 e^{-\sigma}+\cosh (\sigma).
\end{equation}
In this coordinate system, the metric on the surface becomes $d\ell^2=d\sigma^2+e^{-2\sigma}d\mu^2$, which is positively defined.

\begin{figure}
\begin{center}
\includegraphics[width=0.4\textwidth]{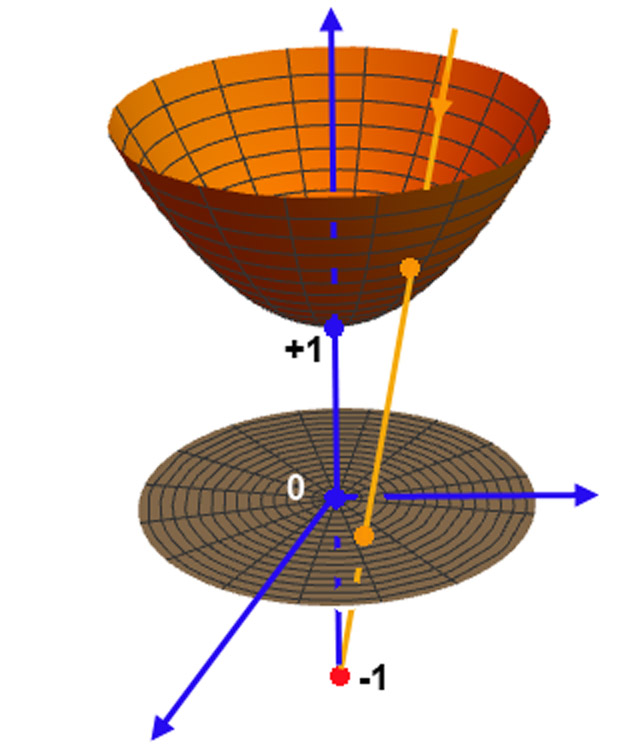}
\end{center}
\caption{Poincar\'e disk representation of the hyperbolic plane $\mathbb{H}^2$.}
\label{fg1}
\end{figure}
%

Besides the above Minkowski hyperboloid model, there are two other well-known conformal representations of $\mathbb{H}^2$, embedded on a usual plane.  We mention these representations here, so that we can use them to visualize $\mathbb{H}^2$ now, and generalize them to 3D later.

First, the Poincar\'e disk is defined by Fig.~\ref{fg1}.  For this representation, we make a stereographic projection from the hyperbolic surface $x_1^2+x_2^2-x_3^2=-1$ onto the horizontal $(x,y)$ plane, by constructing a line from any point on the surface to the pole $(0,0,-1)$.  This line intersects the horizontal plane at the point $x = x_1/(1 + x_3)$, $y = x_2/(1 + x_3)$, which lies inside the unit disk.  Hence, any point inside this disk represents a point on $\mathbb{H}^2$, and any point on the boundary of the disk represents infinity on $\mathbb{H}^2$.  Because of this projection, all distances appear to contract near the boundary.  The Poincar\'e disk representation is conformal but not isometric, the metric reading here $d\ell^2=4(1-x^2-y^2)^{-2}(dx^2+dy^2)$. Geodesic lines are circles (or diameters), orthogonal to the unit circle.

For a second representation, we can transform the Poincar\'e disk into the upper half-plane by a conformal mapping, which is an inversion with a pole located on the boundary of the disk.  The boundary is then mapped onto the $x$ axis, and the disk interior is mapped onto the half plane with positive $y$. Geodesics are then represented by circles (or straight lines) orthogonal to the $x$ axis, and the metric becomes $d\ell^2=y^{-2}(dx^2+ dy^2)$.

\begin{figure}
\begin{center}
\includegraphics[width=0.75\textwidth]{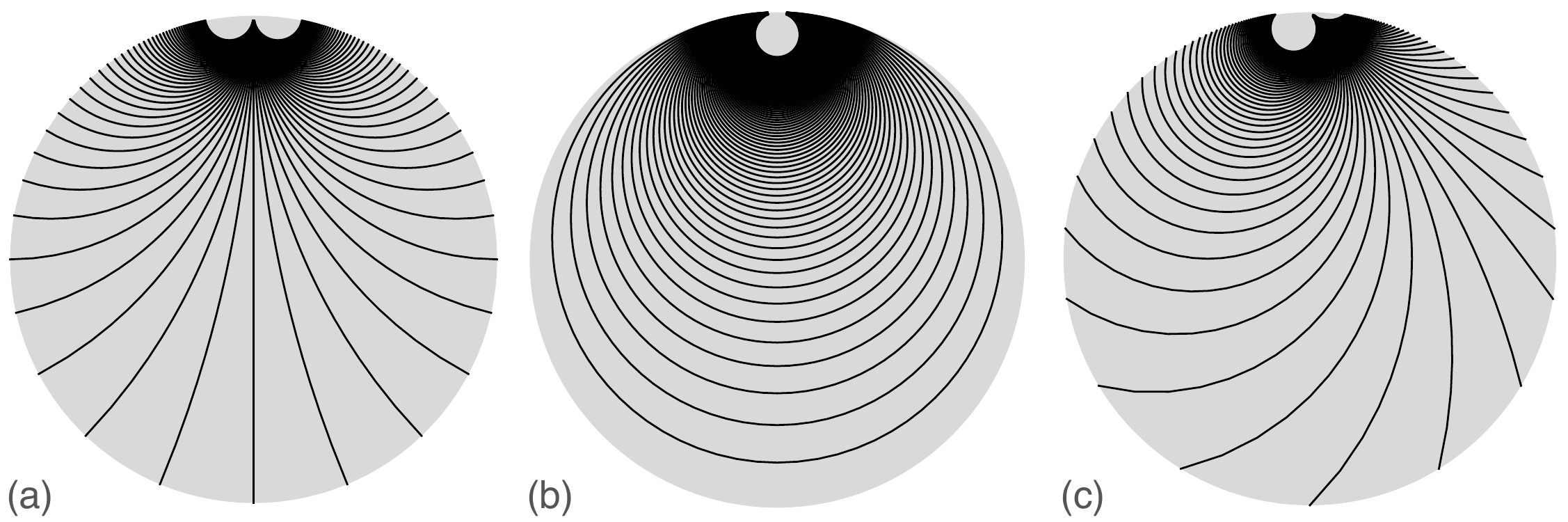} 
\end{center}
\caption{Fields in $\mathbb{H}^2$ derived from the exponential coordinates, represented in the Poincar\'e disk.  (a)~Pure splay field tangent to geodesic lines, obtained by varying $\sigma$ with constant $\mu$ on each line, leading to $S=-1$, $|\mathbf{B}|^2=0$.  (b)~Pure bend field obtained by varying $\mu$ with constant $\sigma$. The field lines are along $\mathbb{H}^2$ horocircles, leading to $S=0$, $|\mathbf{B}|^2=1$. (c)~Field with mixed splay and bend.}
\label{fg2}
\end{figure}
%

We can now define various types of director field in the exponential coordinate system, and visualize them on the Poincar\'e disk.  First, varying $\sigma$ with constant $\mu$ leads to the field lines shown in Fig.~\ref{fg2}a.  These lines are a bundle of parallel geodesics in $\mathbb{H}^2$, which are represented by circular arcs on the Poincar\'e disk.  The corresponding director field $n^i=(1,0)$ has constant splay $S=-1$ and zero bend (as expected because the field lines are geodesics).  Alternatively, varying $\mu$ with constant $\sigma$ leads to the fields lines in Fig.~\ref{fg2}b.  Those lines are called horocircles, and are orthogonal to the previous bundle.  The corresponding director field $n^i=(0,e^\sigma)$ has constant bend $B^i = (-1,0)$ with magnitude $|\mathbf{B}|^2=1$ and zero splay.

For an intermediate case, we consider the director field $n^i=(\cos\theta,e^\sigma \sin\theta)$, as shown in Fig.~\ref{fg2}c.  It has a mixture of splay $S^2=\cos^2 \theta$ and bend $|\mathbf{B}|^2 = \sin^2\theta$, so that $S^2+|\mathbf{B}|^2=1$.  This result is consistent with the compatibility constraint derived by Niv and Efrati~\cite{Niv2018}.

In both Figs.~\ref{fg2}a and~\ref{fg2}b, the field lines converge toward a single point at infinity.  The choice of this point is just an artifact of the coordinate system; any point at infinity could have been chosen.  Hence, there is a continuous set of constant splay fields and a continuous set of constant bend fields, coded by the angular coordinate of the special point at infinity.

The field lines associated with both constant splay and bend cases have a very simple visualization in the upper half-plane representation. If the inversion pole, from the disk to the upper half-plane, is chosen opposite to the point at infinity where geodesics and horocircles meet, then the field lines are mapped to constant $x$ lines (splay case) and constant $y$ lines (bend case).

As an aside, this dual set of geodesic and orthogonal horocircles also plays an interesting role in a different physical problem, which is the Laplace equation in  $\mathbb{H}^2$. Here, the wave-like solutions progress along geodesic bundles (of the pure splay type), with equiphase loci along orthogonal horocircles (of the pure bend type).

From these results, we can see that $\mathbb{H}^2$ is quite different from $\mathbb{R}^2$ and $\mathbb{S}^2$ in an important way:  In $\mathbb{H}^2$, we can construct director fields with pure splay or pure bend of constant magnitude.  In $\mathbb{R}^2$ and $\mathbb{S}^2$, we can construct director fields with pure splay or pure bend, but the magnitude of splay or bend is not uniform, and the fields contain a singularity.  If the Oseen-Frank elastic free energy favors a certain nonzero magnitude of splay or bend, it can achieve the favored state everywhere in $\mathbb{H}^2$.  By contrast, it is frustrated in $\mathbb{R}^2$ and $\mathbb{S}^2$, and cannot achieve the favored state everywhere.

\section{Director fields in 3D spaces of constant curvature}

We now address the case of 3D director fields, and look for geometries in which a single one of the four deformation modes can be realized by itself, without any of the other deformation modes.  For that purpose, we analyze the eight homogeneous 3D spaces of constant scalar curvature \cite{Thurston1997,Scott1983}.  These eight 3D geometries include three simple generalizations of the 2D geometries studied above: $\mathbb{R}^3$, $\mathbb{S}^3$ and $\mathbb{H}^3$.  Those geometries are isotropic and have constant sectional curvature---zero, positive, and negative, respectively.  In addition, we must consider two cases of simple cross products, $\mathbb{S}^2\times\mathbb{R}$ and $\mathbb{H}^2\times\mathbb{R}$, which are anisotropic.  Finally, we have three more complex cases, called $\widetilde{SL}(2,R)$, $Nil$, and $Sol$, which are also anisotropic.

In each of the spaces, we construct director fields, calculate the four deformation modes $S^2$, $|\mathbf{B}|^2$, $T^2$, and $\Tr(\Delta^2)$, and search for configurations in which only one deformation mode is non-vanishing.  We also look empirically for 3D generalizations of the 2D compatibility relation $S^2+|\mathbf{B}|^2=-K_G$.

\subsection{Euclidean space $\mathbb{R}^3$}

Although this is the space of main interest for analyzing real liquid crystals, it does not allow for director fields with a single constant deformation mode---either bend, twist, splay, or biaxial splay---except the trivial constant field for which these deformations all vanish.  Each individual mode can occur \emph{locally} in $\mathbb{R}^3$, as discussed in the review article~\cite{Selinger2018}, but those configurations show nonuniform deformations or even  singularities in the director field.  The allowed combinations of constant deformations have been characterized by Virga~\cite{Virga2019}.

Let us just discuss here the case of twist deformations, expected for liquid crystals with chiral molecules.  Cholesteric phases are often observed in that case.  For example, consider the director field $\hat{\mathbf{n}}(x,y,z)= (\cos qz,\sin qz,0)$.  Applying the framework given in Sec.~2 leads to
$\mathit{\mathbf{B}}=0$, $T=-q$, $S=0$, and
\begin{equation}
\Delta=\frac{q}{2} \left(
\begin{array}{ccc}
 0 & 0 & -\sin(qz) \\
 0 & 0 & \cos qz \\
 - \sin qz & \cos qz & 0 \\
\end{array}
\right).
\end{equation}
Therefore, a cholesteric ``single twist'' configuration can be seen as a mixture of (double) twist $T$ and biaxial splay $\Delta$.

A pure double twist configuration in $\mathbb{R}^3$ can only be realized along a line, as a ``double-twist tube.''  As an example, for a tube along the $z$ axis, the director field is
\begin{equation}
\hat{\mathbf{n}}(x,y,z)=\frac{(-qy,qx,1)}{\sqrt{1+q^2\left(x^2+y^2\right)}},
\end{equation}
Along the $z$ axis, all deformation modes vanish except for (double) twist $T=2q$.  Away from the $z$ axis, this field has a nonuniform mixture of twist and bend.  It can be shown that a pure double twist configuration cannot be defined in $\mathbb{R}^3$, necessarily leading to field singularities.  Indeed, a well-established model for liquid-crystal ``blue phases" is a 3D array of double-twist tubes pierced by a dual array of disclination lines~\cite{Meiboom1983}.

\subsection{Three-dimensional sphere $\mathbb{S}^3$}

The defect-rich structure of the blue phase in $\mathbb{R}^3$ motivated Sethna \emph{et al.}~\cite{Sethna1983} to construct an ideal, defect-free blue phase in the positively curved space $\mathbb{S}^3$.  This structure has pure double twist, of constant magnitude, with none of the other liquid-crystal deformation modes.  Here, we review previous results on the ideal blue phase~\cite{Sethna1983,Pansu1987}, emphasizing the role of Hopf fibrations.

The 3-sphere $\mathbb{S}^3$ is a curved 3D space, which can be embedded in 4D Euclidean space by the simple equation $x_1^2+x_2^2+x_3^2+x_4^2=1$. It can also be put in one-to-one correspondence with the special unitary group $SU(2)$, widely used in physics. Among the different parameterizations of $\mathbb{S}^3$, we shall use here a toroidal coordinate system $(\phi,\theta,\omega)$, which is defined by
\begin{equation}
x_{1}=\cos \theta \sin \phi,\
x_{2}=\sin \theta \sin \phi,\
x_{3}=\cos \omega \cos \phi,\
x_{4}=\sin \omega \cos \phi.
\end{equation}
with  $0\leq\phi\leq\frac{\pi}{2}$, $0\leq\theta\leq2\pi$, and $0\leq\omega\leq2\pi$.  In this coordinate system, the metric is $d\ell^2 = d\phi^2 + \sin^2 \phi\, d\theta^2 + \cos^2 \phi\, d\omega^2$. The $\mathbb{S}^3$ scalar curvature reads $R=6$, and its  isotropic nature is seen in $diag({R^i}_j)=(2,2,2)$.

%
\begin{figure}
\begin{center}
\includegraphics[width=0.5\textwidth]{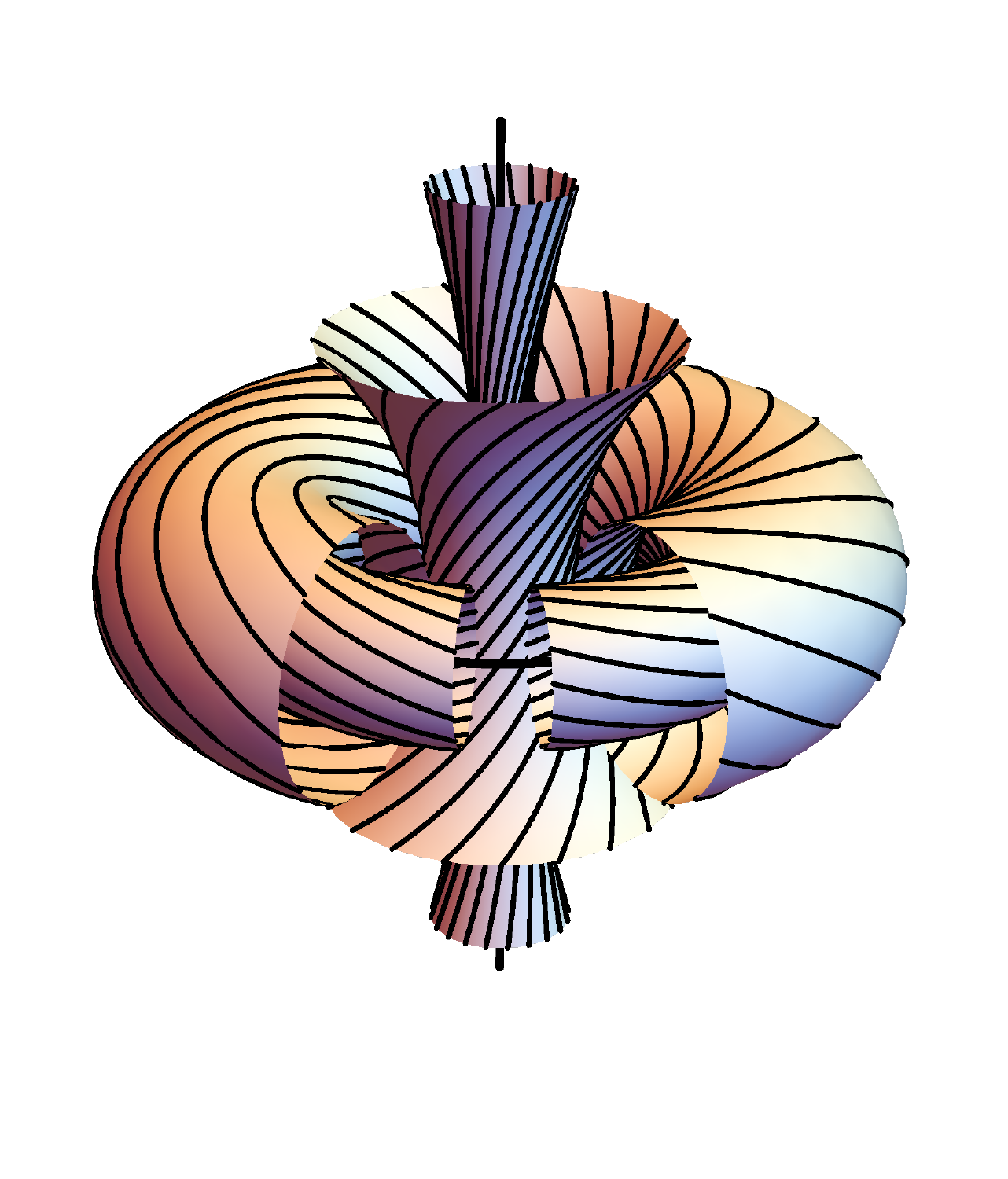}
\end{center}
\caption{ Stereographic projection onto $\mathbb{R}^3$ of a part of the $\mathrm{S}^3$ great circles Hopf fibration.  These circles are on a torus bundle, some of which being drawn here, organized around two  interlaced great circles of $\mathbb{S}^3$.}
\label{fg4}
\end{figure}
%

Surfaces of constant $\phi$ are tori, and these tori form a foliation of $\mathbb{S}^3$; i.e.\ they break the space into a series of sheets.  Figure~\ref{fg4} shows a stereographic projection of these tori onto $\mathbb{R}^3$.  The tori are organised around two interlaced great circles $x_{1}^2 + x_{2}^2 =1$ and $x_{3}^2 + x_{4}^2 =1$, corresponding to $\phi=\frac{\pi}{2}$  and $\phi=0$, which act as two orthogonal $C_\infty$ symmetry axes of $\mathbb{S}^3$.  In the stereographic projection, one of those great circles maps onto the horizontal circle at the core of the torus, and the other maps onto the vertical axis. 

Consider a specific torus in the foliation, with the coordinate $\phi=\phi_0$.  Within that torus, we impose a simple linear relation between the coordinates $\theta$ and $\omega$, of the form $\theta = \omega + \omega_0$, with $\omega_0$ constant.  This relation defines a curve on the torus, which is a great circle.  Varying the parameter $\omega_0$ leads to a fibration of the torus; i.e.\ it breaks the torus into a series of fibres, which are all great circles.  These fibres are indicated by black lines on each torus in Fig.~\ref{fg4}.  If we vary both the parameters $\phi_0$ and $\omega_0$, we obtain a great circle fibration of the entire space $\mathbb{S}^3$; i.e.\ it breaks the space into fibres, with each circular fibre associated with the pair $(\phi_0,\omega_0)$.  This construction is called the Hopf fibration.  Note that the two symmetry axis circles,  with $\phi=0$ and $\frac{\pi}{2}$, are part of this fibration.

Following the previous work, we define a director field $\hat{\mathbf{n}}(\phi,\theta,\omega)$ in $\mathbb{S}^3$ to be everywhere tangent to the Hopf fibration.  This director field has contravariant components $n^i=(0,1,1)$, covariant components $n_i=(0,\sin^2\phi,\cos^2\phi)$, and hence is normalized.  By putting this director field into Eqs.~(\ref{equa4}--\ref{equa7}), we can calculate all four of the deformation modes.  Explicit calculations give double twist $T=2$, splay $S=0$, bend $\mathbf{B}=0$, and biaxial splay $\Delta_{ij}=0$.  (The zero bend is expected, because the director field is tangent to great circles, which are geodesics in $\mathbb{S}^3$.)  Hence, this structure has pure double twist, independent of position, with none of the other deformation modes.

If we repeat the calculation in a 3-sphere of radius $\rho$, we obtain a double twist of $T=2/\rho$, again with zero splay, bend, and biaxial splay.  Hence, if a chiral liquid crystal has the free energy of Eq.~(\ref{equa9}), which favors $T=2q$ and all other modes zero, it can reach an ideal, unfrustrated state in a 3-sphere of radius $\rho=1/q$.  By contrast, in a 3-sphere of any other curvature radius (or in the limiting case of $\mathbb{R}^3$ with infinite radius), the structure is frustrated, meaning that it is unable to fill up space with the ideal local deformation.  In that case, it must form a blue phase with director singularities, or a cholesteric phase with a combination of double twist and biaxial splay.

As a further point, we note that, due to the isotropy of $\mathbb{S}^3$, there is a continuum of Hopf fibrations, obtained from the given fibration by applying symmetry elements of $\mathbb{S}^3$.  It is also possible to generate Hopf fibrations with the opposite chirality by applying a reflection.

Another interesting director configuration in $\mathbb{S}^3$ is tangent to field lines with varying $\phi$ and constant $\theta$ and $\omega$.  This director field $n^i=(1,0,0)$ is normal to the tori.  The field lines are geodesic, so there is still no bend $\mathbf{B}=0$.  Furthermore, there is no twist $T=0$, and we are left with splay $S=2\cot2\phi$ and biaxial splay $\Tr(\Delta^2)=2\csc^2 2\phi$.

\subsection{Hyperbolic space $\mathbb{H}^3$}

To fulfill our program of finding ideal director fields for each liquid-crystal deformation mode, we now consider the six other 3D homogeneous spaces, starting with the hyperbolic space $\mathbb{H}^3$.

To analyze the 3D hyperbolic space $\mathbb{H}^3$, we follow a procedure analogous to the 2D hyperbolic plane $\mathbb{H}^2$ in Sec.~3.3.  The space $\mathbb{H}^3$ can be represented as the hyperboloid $x_1^2+x_2^2+x_3^2-x_4^2=-1$ embedded in 4D Minkowski space, provided that the 4D metric is defined as $d\ell^2=dx_1^2+dx_2^2+dx_3^2-dx_4^2$.  To parameterize $\mathbb{H}^3$, we choose the exponential coordinate system $(\mu,\nu,\sigma)$ defined by~\cite{Costa2001}
\begin{eqnarray}
x_1&=&\mu  e^{-\sigma },\quad
x_2=\nu  e^{-\sigma },\quad
x_3=\frac{1}{2} e^{-\sigma } \left(\mu ^2+\nu ^2\right)+\sinh (\sigma ),\nonumber\\
x_4&=&\frac{1}{2} e^{-\sigma } \left(\mu ^2+\nu ^2\right)+\cosh (\sigma ).
\end{eqnarray}
In this coordinate system, the positively defined metric reads $d\ell^2=e^{-2\sigma}(d\mu^2+d\nu^2)+d\sigma^2$. The $\mathbb{H}^3$ scalar curvature is $R=-6$, and its isotropic nature is seen in $diag({R^i}_j)=(-2,-2,-2)$.

%
\begin{figure}
\begin{center}
\includegraphics[width=0.5\textwidth]{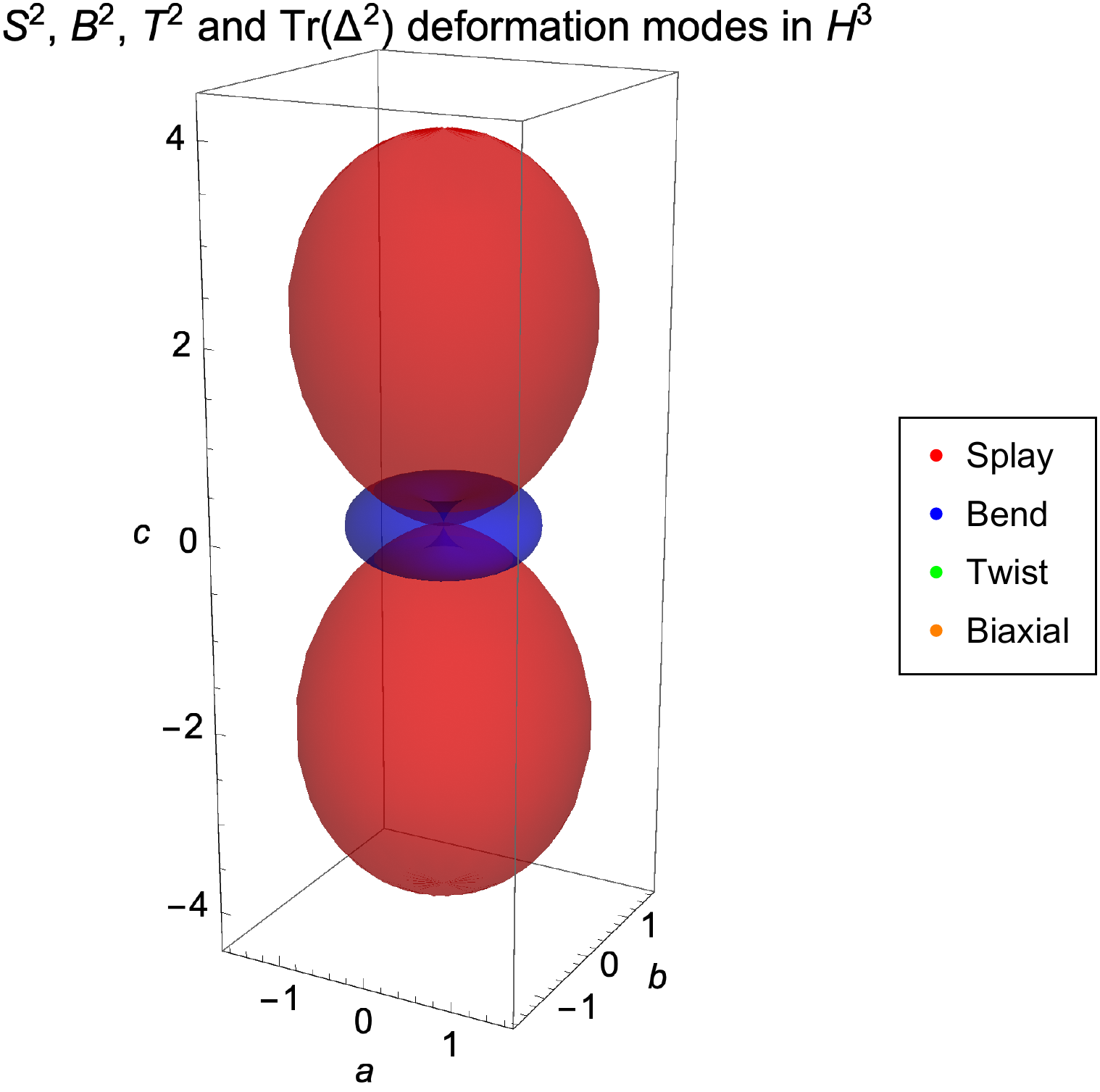}
\end{center}
\caption{Deformation modes for the $\mathbb{H}^3$ manifold. The director field has a fixed orientation, given by the angles $\alpha$ and $\beta$ in the $\{\hat{\mathbf{a}},\hat{\mathbf{b}},\hat{\mathbf{c}}\}$ basis.  The intensity of each deformation mode is represented by the distance of the corresponding surface from the origin, at that orientation of $\hat{\mathbf{n}}$.  The twist and biaxial splay both vanish.}
\label{chemicalH3}
\end{figure}
%

Because the eigenvalues of the mixed Ricci tensor are all degenerate, this tensor is proportional to the identity, and all vectors are eigenvectors.  For that reason, we have complete freedom to choose any three orthonormal vectors as our set of Ricci eigenvectors.  One convenient choice is just along the conventional tangent vectors, so that $a^i=(e^\sigma,0,0)$, $b^i=(0,e^\sigma,0)$, and $c^i=(0,0,1)$.  Following Eq.~(\ref{eqn:field}), we define the director $\hat{\mathbf{n}}$ in terms of this basis as $\hat{\mathbf{n}}=\sin \beta \cos \alpha \,\hat{\mathbf{a}} +\sin \beta \sin \alpha\,\hat{\mathbf{b}}+ \cos \beta\,\hat{\mathbf{c}}$, with constant azimuthal angle $\alpha$ and polar angle $\beta$.  A direct computation of the different deformation modes then gives
\begin{equation}
\label{paramH3}
S^2= 4 \cos^2 \beta, \quad |\mathbf{B}|^2= \sin^2 \beta,\quad T^2= 0, \quad \Tr(\Delta^2)=0.
\end{equation}
All of the director fields in this family satisfy the relation
\begin{equation}
S^2+ 4 |\mathbf{B}|^2=4.
\label{CH3}
\end{equation}
Figure~\ref{chemicalH3} shows a polar plot of the four deformation modes as functions of the angles $\alpha$ and $\beta$.  Two special cases are particularly interesting:  The director field has pure constant splay if $\beta=0$, and it has pure constant bend if $\beta=\pi/2$.  In the latter case, note that there is a continuous set of pure bend field upon varying $\alpha$.

%
\begin{figure}
\begin{center}
\includegraphics[width=0.5\textwidth]{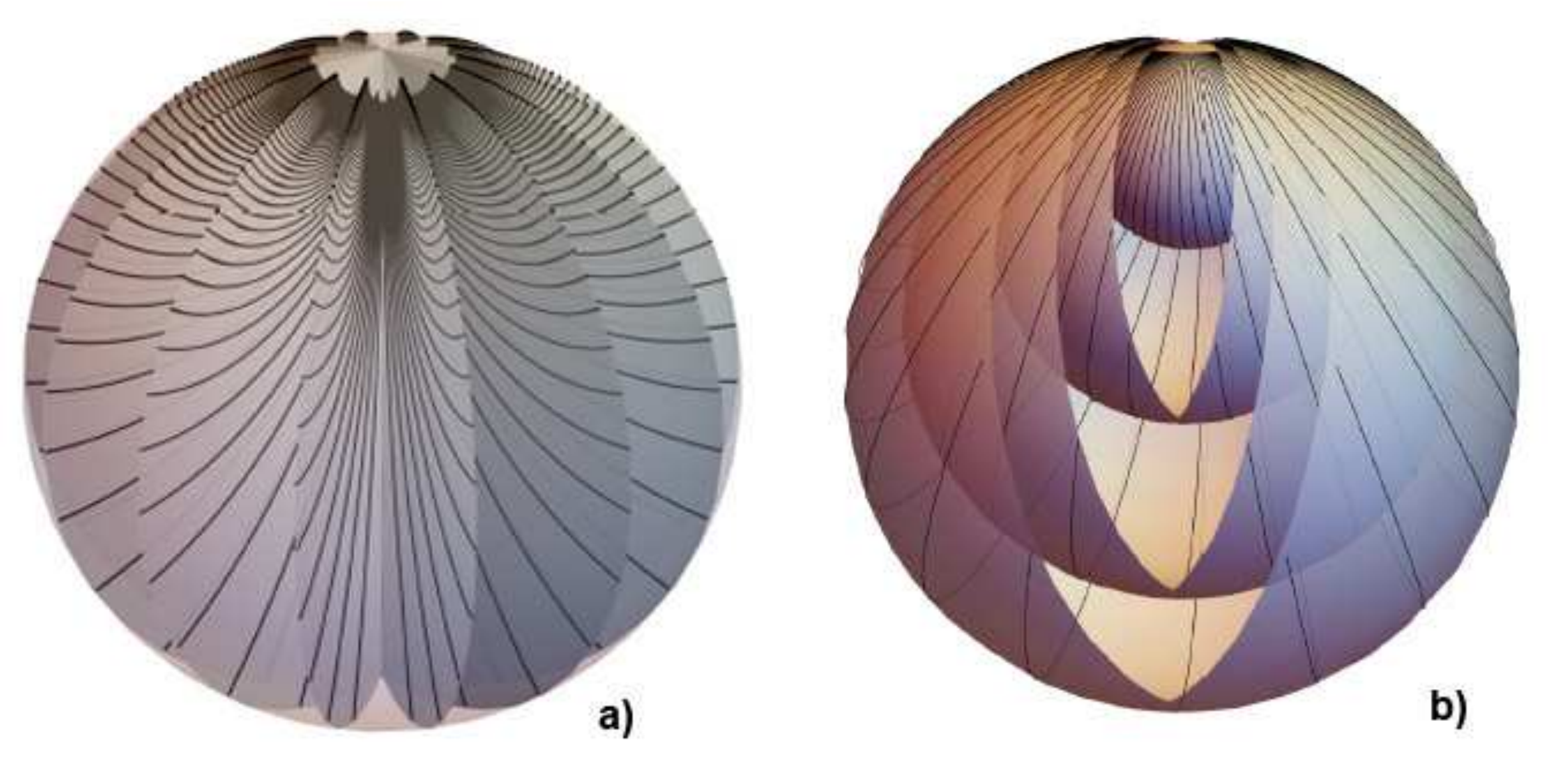}
\end{center}
\caption{Pure splay and bend configurations in the Poincar\'e ball representation of $\mathbb{H}^3$. (a) Pure splay field lines, along parallel geodesics coming from a point at infinity (here the north pole on the limit sphere). (b) Pure bend field lines, tangent to a bundle of horocircles drawn on a bundle of horospheres (here tangent to the north pole on the limit sphere).}
\label{fg6}
\end{figure}
%

To visualize these director fields in $\mathbb{H}^3$, we use the Poincar\'e ball representation, analogous to the Poincar\'e disk discussed in Sec.~3.3.  For this representation, we make a stereographic projection from the hyperboloid onto the $(x,y,z)$ space, by constructing a line from any point in the hyperboloid to the pole $(0,0,0,-1)$.  This line maps any point of the hyperboloid $(x_1,x_2,x_3,x_4)$ onto $x=x_1/(1+x_4)$, $y=x_2/(1+x_4)$, $z=x_3/(1+x_4)$, which lies inside the unit ball $B^3$.  In this Poincar\'e ball, points at infinity of $\mathbb{H}^3$ are represented by the limit sphere $R=1$.  Geodesics are represented by arcs of circles orthogonal to the limit sphere, as shown in Fig.~\ref{fg6}a, and geodesic surfaces by spherical caps orthogonal to the limit sphere.  Horospheres are spheres tangent to the limit sphere, shown in Fig.~\ref{fg6}b, are horocircles are fibres on those spheres.

Now the the two special cases of director field are easily identified. The pure splay field (with $\beta=0$) amounts to  varying $\sigma$ with constant $\mu$ and $\nu$, leading to the geodesic field lines in Fig.~\ref{fg6}a, which converge toward a single point at infinity.  The corresponding director field $\hat{\mathbf{n}}=\hat{\mathbf{c}}$ has pure constant splay, with zero bend, twist, and biaxial splay.  (The zero bend is expected because the field lines are geodesics.)  There is a continuous set of such director fields with pure constant splay, which can be obtained by global rotations changing the bundle end point on the unit sphere bounding the Poincar\'e ball.

For the pure bend field (with $\beta=\pi/2$), we can use any unit vector in the $\{\hat{\mathbf{a}},\hat{\mathbf{b}}\}$ plane. For example,  varying $\mu$ with constant $\sigma$ and $\nu$ leads to the field lines shown in Fig.~\ref{fg6}b.  Here, the constant $\sigma$ defines a horosphere, and the constant $\nu$ selects a horocircle on that horosphere.  We can therefore build an overall horocircle bundle of $\mathbb{H}^3$.  The corresponding director field tangent to these horocircles is $\hat{\mathbf{n}}=\hat{\mathbf{a}}$, which has pure constant bend, along with zero splay, twist, and biaxial splay.

Note that the pure bend field lines are everywhere orthogonal to the pure splay field lines.  Therefore, the chosen exponential coordinate system gives a local basis of tangent vectors with pure constant splay along one direction, and pure constant bend along the other two.

From these results, we can see that the hyperbolic space $\mathbb{H}^3$ allows director fields with pure constant splay or with pure constant bend, just as the 3-sphere $\mathbb{S}^3$ allows director fields with pure constant twist.  If we change the curvature radius of $\mathbb{H}^3$ from $1$ to $\rho$, the splay of the first director field would change to $S^2=4/\rho^2$, and the bend of the second director field would change to $|\mathbf{B}|^2=1/\rho^2$.  Hence, if the free energy for bent-core molecules favors a certain bend, as in Eq.~(\ref{equa10}), or the free energy for pear-shaped molecules favors a certain splay, as in Eq.(\ref{equa11}), the ideal structure can be achieved without frustration in a hyperbolic space with the appropriate curvature radius.

As previously discussed for $\mathbb{H}^2$, we can also transform, by inversion, from a Poincar\'e ball visualization to an upper half-space visualization of $\mathbb{H}^3$.  This upper half-space visualization is defined for positive $z$, and it has the metric $d\ell^2=z^{-2}(dx^2+ dy^2+dz^2)$.  In this visualization, the pure splay case corresponds to field lines along $z$, and the pure bend case corresponds to parallel lines in the $(x,y)$ plane.

We now extend the analysis to the remaining five 3D homogeneous geometries, which are anisotropic, especially to search for an example of pure constant biaxial splay.

\subsection{Space $\mathbb{S}^2\times\mathbb{R}$}

The space $\mathbb{S}^2\times\mathbb{R}$ consists of the 2D sphere $\mathbb{S}^2$ extended uniformly in an orthogonal direction.  A natural coordinate system $(\theta,\phi,z)$ is just the $(\theta,\phi)$ angular coordinates of $\mathbb{S}^2$ combined with the $z$ coordinate along $\mathbb{R}$.  In this coordinate system, the metric is $d\ell^2= d\theta^2+\sin^2 \theta\, d\phi^2+dz^2$.  This space is clearly anisotropic, because the $z$ direction is different from the other two directions.  The $(\theta,\phi)$ surface is a sphere with positive Gaussian curvature, while any surface containing $z$ is a cylinder with zero Gaussian curvature. The scalar curvature is $R=2$, and its anisotropic nature is seen in $diag({R^i}_j)=(1,1,0)$.

One trivial director field in this space is just in the $z$ direction, with $n^i=(0,0,1)$.  This director field has zero splay, twist, bend, and biaxial splay.

Two other possibilities are the director fields on the 2D sphere $\mathbb{S}^2$ defined in Sec.~3.2, with no component in the $z$ direction.  Along the meridians, varying $\theta$ with constant $\phi$ and $z$ gives the director field $n^i=(1,0,0)$.  It has nonuniform splay $S^2=\cot^2 \theta$ and nonuniform biaxial splay $\Tr(\Delta^2)=\frac{1}{2}\cot^2 \theta$, with zero twist and zero bend.  Along the parallel circles, varying $\phi$ with constant $\theta$ and $z$ gives the director field $n^i=(0,\csc\theta,0)$.  It has nonuniform bend $|\mathbf{B}|^2=\cot^2 \theta$.

In each of the latter two cases, the director field has singularities at $\theta=0$ and $\pi$.  To avoid the free energy cost of these singularities, the director field might escape into the third dimension by developing a nonzero $z$ component.  In that sense, it could form a non-Euclidean analogue of the structure that is often observed in liquid-crystal tubes, which has bend, splay, and biaxial splay.

The space $\mathbb{S}^2\times\mathbb{R}$ is apparently not useful for our current purpose of finding geometries with pure constant deformation modes.  However, as an aside, we should note that this space was previously studied to solve another soft-matter frustration problem, which was in lamellar and smectic liquid crystals~\cite{Sadoc1999,Charvolin1987,Charvolin1988}.  In that case, the special feature of $\mathbb{S}^2\times\mathbb{R}$ geometry is to provide equal spacing between smectic layers.

\subsection{Space $\mathbb{H}^2\times\mathbb{R}$}

The space $\mathbb{H}^2\times\mathbb{R}$ is the 2D hyperbolic plane $\mathbb{H}^2$ extended uniformly in an orthogonal direction.  We construct a coordinate system $(\sigma,\mu,z)$ by combining the exponential coordinates $(\sigma,\mu)$ of $\mathbb{H}^2$, which were presented in Sec.~3.3, with the $z$ coordinate for $\mathbb{R}$.  In this coordinate system, the metric is $d\ell^2=d\sigma^2+e^{-2\sigma}d\mu^2+dz^2$. The scalar curvature is $R=-2$, and its  anisotropic nature is seen in $diag({R^i}_j)=(-1,-1,0)$.

%
\begin{figure}
\begin{center}
\includegraphics[width=0.5\textwidth]{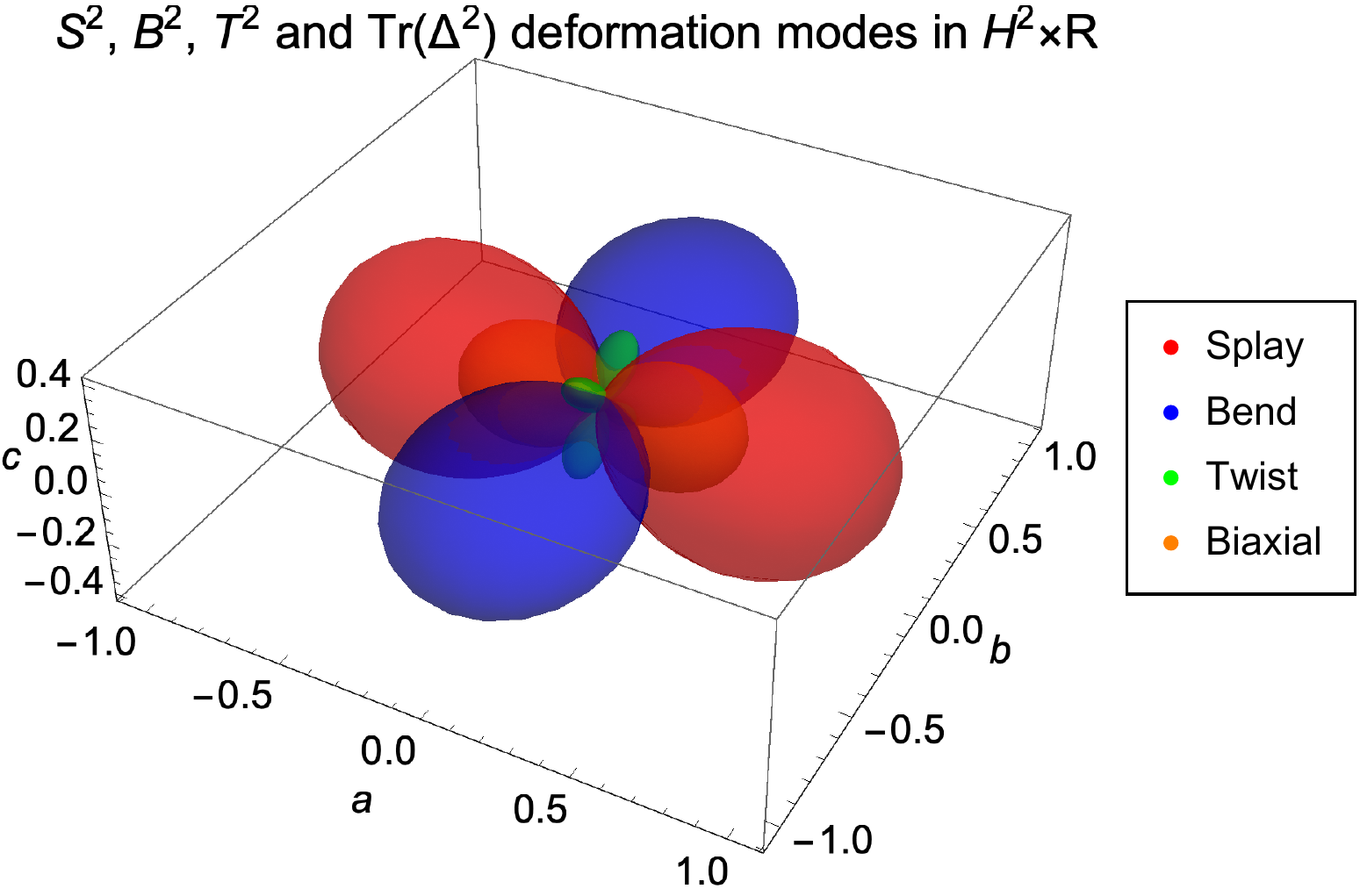}
\end{center}
\caption{Deformation modes for the $\mathbb{H}^2\times\mathbb{R}$ manifold.  The intensity of each deformation mode is represented by the distance of the corresponding surface from the origin, at that orientation of $\hat{\mathbf{n}}$.}
\label{chemicalH2R}
\end{figure}
%

Because the third eigenvalue of the mixed Ricci tensor is distinct from the other two, the corresponding eigenvector must be $c^i=(0,0,1)$.  The other two orthonormal eigenvectors can be chosen as $a^i=(1,0,0)$ and $b^i=(0,e^\sigma,0)$.  Following Eq.~(\ref{eqn:field}), we define the director $\hat{\mathbf{n}}$ with respect to this basis as $\hat{\mathbf{n}}=\sin \beta \cos \alpha \,\hat{\mathbf{a}} +\sin \beta \sin \alpha\,\hat{\mathbf{b}}+ \cos \beta\,\hat{\mathbf{c}}$, with constant $\alpha$ and $\beta$.  Direct computation of the deformation modes then gives
\begin{eqnarray}
S^2=\cos^2 \alpha\sin^2 \beta,\quad |\mathbf{B}|^2=\sin^2 \alpha\sin^4 \beta,\quad T^2=\sin^2 \alpha\sin^2 \beta\cos^2 \beta,\nonumber\\ \Tr(\Delta^2)=\frac{1}{2}\left[\cos^2 \alpha\sin^2 \beta+\sin^2 \alpha\sin^2 \beta\cos^2 \beta\right]
=\frac{1}{2} \left[S^2+T^2\right].
\label{paramH2R}
\end{eqnarray}
Figure~\ref{chemicalH2R} shows a polar plot of the four deformation modes as functions of the angles $\alpha$ and $\beta$.  In these expressions, we do not see a general relation analogous to Eq.~(\ref{CH3}) that is valid in the whole space.  However, $\mathbb{H}^2\times\mathbb{R}$ has a natural product structure, suggesting that we may look for an expression value in the $\mathbb{H}^2$ part, where $\beta=\pi/2$.  Here, we find the relation
\begin{equation}
S^2+2|\mathbf{B}|^2+T^2+2\Tr(\Delta^2)=2.
\label{CH2R}
\end{equation}
Note that we could have omitted the $T^2$ term because it vanishes on the plane $\beta=\pi/2$.  However, we keep it for consistency with other relations that will be found on 2D eigenplanes of the mixed Ricci tensor in other geometries, discussed below.

%
\begin{figure}
\begin{center}
(a)\includegraphics[height=0.5\textwidth]{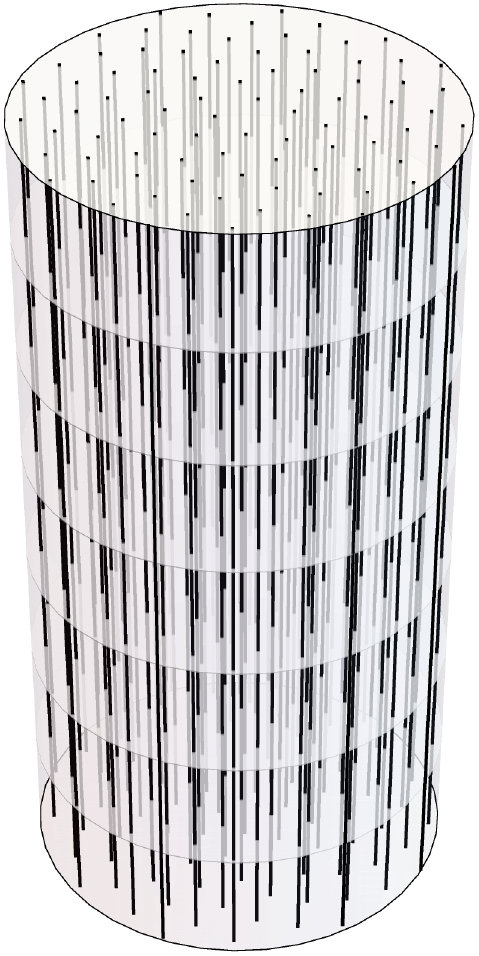}
(b)\includegraphics[height=0.5\textwidth]{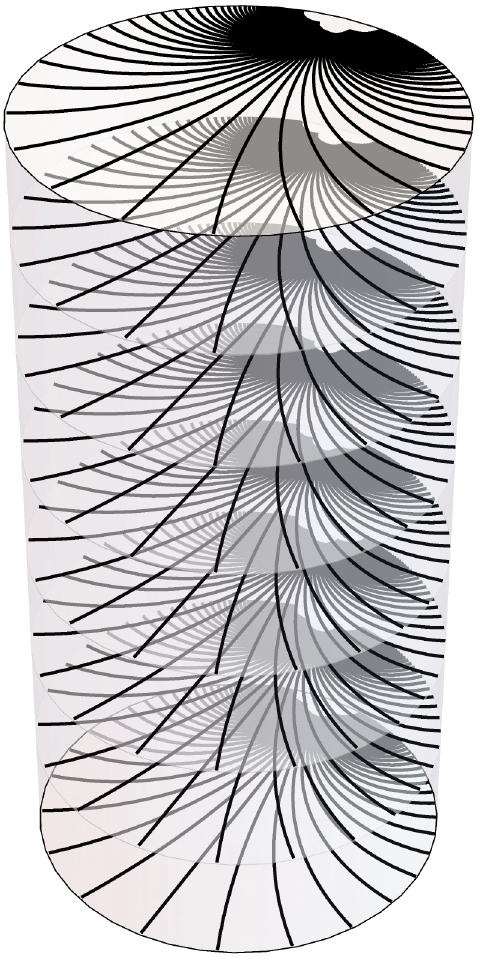}
(c)\includegraphics[height=0.5\textwidth]{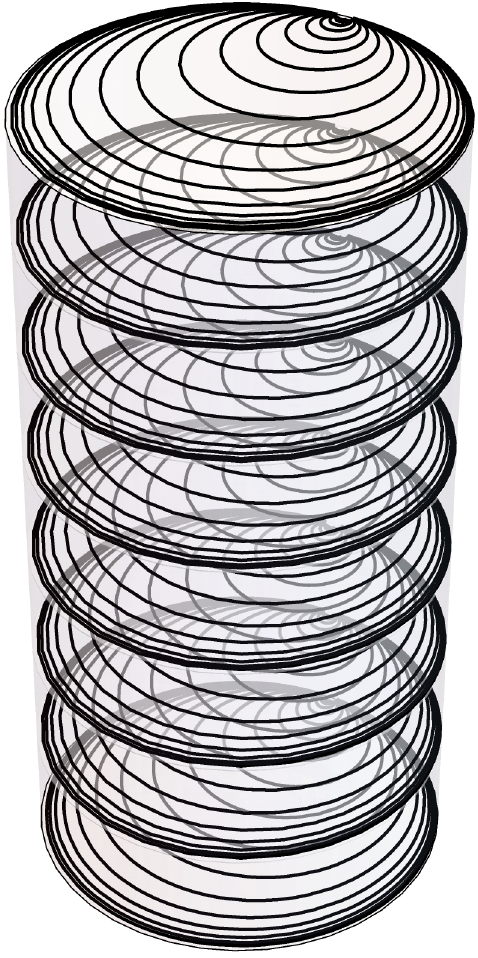}
\end{center}
\caption{Field lines in $\mathbb{H}^2\times\mathbb{R}$ represented in 3D Euclidean space with a horizontal Poincar\'e disk for $\mathbb{H}^2$ and an orthogonal $\mathbb{R}$ vertical direction. (a) Trivial uniform field in the $z$ direction, with zero director gradients. (b) The $\mathbb{H}^2$ geodesic pure splay line bundle, repeated by translation along the $z$ direction.  This director field has constant splay $S$ and biaxial splay $\Delta$, with zero bend and zero twist.  (c) The $\mathbb{H}^2$ pure bend horocircles bundle, repeated by translation along the $z$ direction.  This director field shows constant bend and vanishing splay, twist, and biaxial splay.}
\label{fg8}
\end{figure}
%

To visualize director fields in this space, we extend the Poincar\'e disk into a tube along the vertical axis, as shown in Fig.~\ref{fg8}.  A first, trivial director field is just $\hat{\mathbf{n}}=\hat{\mathbf{c}}$, i.e.\ $\beta=0$, represented in Fig.~\ref{fg8}a.  This director field is parallel to $\mathbb{R}$ in the product manifold $\mathbb{H}^2\times\mathbb{R}$, which is vertical in this visualization.  It is nematic-like with vanishing splay, twist, bend, and biaxial splay.

For more interesting director fields, one may extend the 2D fields found in $\mathbb{H}^2$ to $\mathbb{H}^2\times\mathbb{R}$. This amounts to fixing $\beta=\pi/2$.  As a first possibility, we begin with the pure splay director field on $\mathbb{H}^2$, with the director tangent to geodesic bundles.  In $\mathbb{H}^2\times\mathbb{R}$, this director field is parallel to $\hat{\mathbf{a}}$, corresponding to $\alpha=0$, represented in Fig.~\ref{fg8}b.  It has constant splay $S^2=1$ and constant biaxial splay $\Tr(\Delta^2)=\frac{1}{2}$, with zero bend and zero twist.  We emphasize that this field is not pure splay in 3D, because it includes a component of $\Delta$.  It is an example of planar splay, which is mixed splay and $\Delta$, as discussed in the review article~\cite{Selinger2018}.

As an alternative, we can begin with the pure bend director field on $\mathbb{H}^2$, with the director orthogonal to the geodesic bundles.  In $\mathbb{H}^2\times\mathbb{R}$, this director field is parallel to $\hat{\mathbf{b}}$, corresponding to $\alpha=\pi/2$, represented in Fig.~\ref{fg8}c.  It has constant bend $|\mathbf{B}|^2=1$, and zero splay, twist, and biaxial splay.  We emphasize that this field is pure bend in 3D.  The difference between this case with pure bend and the previous case with mixed splay and $\Delta$ arises from a fundamental distinction between bend and splay in 3D:  Pure bend is a planar deformation, but pure splay is double splay, which is not a planar deformation.  That distinction occurs in 3D but not in 2D, where bend and splay are both planar deformations.

\subsection{The $\widetilde{SL}(2,R)$ geometry}

The next Thurston geometry, $\widetilde{SL}(2,R)$, is related to $SL(2,R)$, the group of  $2\times 2$ matrices with real entries and unit determinant. At a topological level, $\widetilde{SL}(2,R)$ is the ''universal cover'' of $SL(2,R)$, meaning that the latter is not simply connected. The simplest example of a non-simply connected manifold is the unit circle $\mathbb{S}^1$, because a path encircling the origin cannot be deformed continuously to a non-encircling path. The covering space of $\mathbb{S}^1$ is $\mathbb{R}^1$, derived by shifting the $2\pi$-periodic coordinate to the full real line. A similar process is followed from $SL(2,R)$ to $\widetilde{SL}(2,R)$.

To see how $SL(2,R)$ arises in the context of 3D manifolds, let us recall that $\mathbb{H}^3$ was above described in $\mathbb{R}^4$ with a Minkowski metric $\{1,1,1,-1\}$, as a sheet of the 3D hyperboloid defined by the equation  $x_1^2+x_2^2+x_3^2-x_4^2=-1$.
This construction can be generalized \cite{Hasebe2010} to other hyperboloids $H^{p,q}$ with
$\sum_{i=1}^p x_i^2-\sum_{j=p+1}^{p+q+1}x_j^2=-1$. With these notations, the hyperboloid used to represent $\mathbb{H}^3$ reads $H^{3,0}$, and even $\mathbb{S}^3$ could be defined as $H^{0,3}$.

The so-called anti-de-Sitter  space $AdS_3$ corresponds to $H^{2,1}$ with equation $x_1^2+x_2^2-x_3^2-x_4^2=-1$, and with a Minkowski signed metric of the form $\{1,1,-1,-1\}$. 

Any element $M \in SL(2,R)$ can be written as
\begin{equation}
M=\left(
\begin{array}{cc}
 x_1+x_4 & x_2+x_3  \\
x_2-x_3 & x_4-x_1 \\
\end{array}
\right)
\end{equation}
with $\det M= -x_1^2-x_2^2+x_3^2+x_4^2=1$.  Hence, there is a one-to-one correspondence between $AdS_3$ and $SL(2,R)$, and a 3D manifold can be associated with this group. We could now follow a similar analysis as in the $\mathbb{H}^3$ case, and define three coordinates on the generalized hyperboloids to search for interesting vector fields. However, this procedure is not convenient here because the induced metric on the hyperboloid, inherited from the Minkowski metric, is generically not positive definite (leading to vectors having possibly negative norm).

We shall therefore proceed differently, and follow Ref.~\cite{Koike1994}, where $\widetilde{SL}(2,R)$ is described as the covering space of the unit tangent bundle of $\mathbb{H}^2$, denoted $U \mathbb{H}^2$, the latter being also closely related to $SL(2,R)$. A unit vector at a point $(x,y)$ of $\mathbb{H}^2$ (say given in the upper plane model) needs another quantity $z$, in the range from $0$ to $2\pi$, to capture the local orientation of that unit vector. Lifting from $SL(2,R)$ to $\widetilde{SL}(2,R)$ amounts to letting the $z$ coordinate run along $\mathbb{R}$. Although this representation of $\widetilde{SL}(2,R)$ topologically recalls that of  $\mathbb{H}^2\times\mathbb{R}$, the two geometries are different, as demonstrated by the explicit form of their metric and Ricci tensor. 

With the above defined $(x,y,z$) coordinates, the covariant metric tensor is \cite{Koike1994}
\begin{equation}
g_{ij}= \frac{1}{y^2}\left(
\begin{array}{ccc}
2 & 0 & y \\
0 & 1 & 0 \\
y & 0 & y^2 \\
\end{array}
\right),
\end{equation}
and the Ricci tensor $R_{ij}$ becomes
\begin{equation}
R_{ij}=\frac{1}{2y^2}\left(
\begin{array}{ccc}
-2 & 0 & y \\
 0 & -3 & 0 \\
 y & 0 & y^2 \\
\end{array}
\right).
\end{equation}
The scalar curvature is $R=-\frac{5}{2}$ for this negatively curved space, with an anisotropy shown by $diag({R^i}_j)=(-\frac{3}{2},-\frac{3}{2},\frac{1}{2})$.

%
\begin{figure}
\begin{center}
\includegraphics[width=0.5\textwidth]{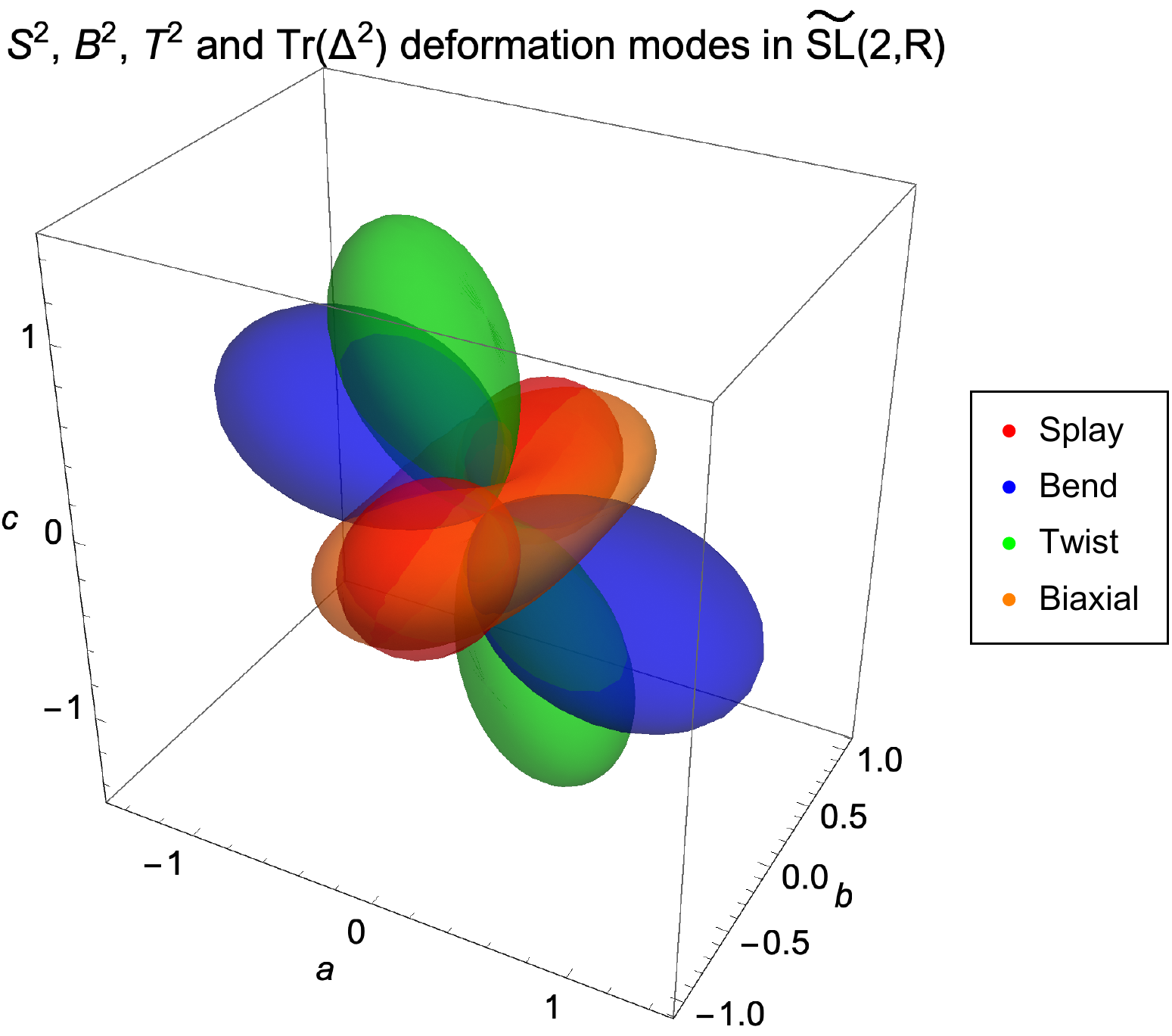}
\end{center}
\caption{Deformation modes for the $\widetilde{SL}(2,R)$ manifold.  The intensity of each deformation mode is represented by the distance of the corresponding surface from the origin, at that orientation of $\hat{\mathbf{n}}$.}
\label{chemicalsl2r}
\end{figure}
%

Because the third eigenvalue of the mixed Ricci tensor is distinct from the other two, the corresponding eigenvector must be $c^i=(0,0,1)$.  The other two orthonormal eigenvectors can be chosen as $a^i=(-y,0,1)$ and $b^i=(0,y,0)$.  Again, we define the director $\hat{\mathbf{n}}$ in this basis as $\hat{\mathbf{n}}=\sin \beta \cos \alpha \,\hat{\mathbf{a}} +\sin \beta \sin \alpha\,\hat{\mathbf{b}}+ \cos \beta\,\hat{\mathbf{c}}$, with constant $\alpha$ and $\beta$.  Direct computation then gives the deformation modes
\begin{eqnarray}
S^2 = \sin^2 \alpha\sin^2 \beta,\quad
|\mathbf{B}|^2 = \sin^2 \beta[\cos\beta-\cos\alpha\sin\beta]^2,\nonumber\\
T^2 = \cos^2 \beta[\cos\beta-\cos\alpha\sin\beta]^2,\nonumber\\ 
\Tr(\Delta^2) = \frac{1}{8}\sin^2\beta\left[5-\cos2\beta-2\cos2\alpha\sin^2 \beta+4\cos\alpha\sin2\beta\right].
\end{eqnarray}
Figure~\ref{chemicalsl2r} shows a polar plot of the four deformation modes.  In these expressions, we cannot find a general relation that is valid in the whole space.  However, following the same reasoning as in the $\mathbb{H}^2\times\mathbb{R}$ case, we focus on the mixed Ricci eigenplane spanned by $\hat{\mathbf{a}}$ and $\hat{\mathbf{b}}$, where $\beta=\pi/2$.  In that plane, we find
\begin{equation}
S^2+2|\mathbf{B}|^2+T^2+2\Tr(\Delta^2)=3.
\label{CSL2R}
\end{equation}
This relation is similar to Eq.~(\ref{CH2R}) for $\mathbb{H}^2\times\mathbb{R}$, except that the constant is now 3 rather than 2.

These results allow us investigate vector fields defined on this manifold, and we find two interesting cases. 

The first case is $\hat{\mathbf{n}}=\hat{\mathbf{c}}$, i.e.\ $\beta=0$.  This director field follows field lines in the direction of variable $z$, leading to a pure constant twist configuration characterized by $T^2=1$.  This result might be surprising, because it shows that global positive curvature, as in the Hopf field on $\mathbb{S}^3$, is not a necessary condition for having constant twist.  Note however that, because $\mathbb{S}^3$ is isotropic, pure twist solutions can be defined in any direction in the tangent space.  By contrast, because $SL(2,R)$ is anisotropic, the pure twist field can only occur along one direction, which is the direction along which the average sectional curvature is positive.  We therefore conjecture that a positive average sectional curvature along one direction might be a necessary condition for the existence of a pure twist field configuration. This point will be discussed again in the $Nil$ space case.

The second interesting case is provided by $\hat{\mathbf{n}}=(\hat{\mathbf{a}}+\hat{\mathbf{c}})/\sqrt{2}$, i.e.\ $\alpha=0$ and $\beta=\pi/4$, so that $n^i=(-y,0,1)/\sqrt{2}$.  Explicit calculation shows that this vector field has zero splay, zero twist, zero bend, but constant nonzero biaxial splay with $\Tr(\Delta^2)=1/2$.  The tensor structure of the biaxial splay is
\begin{equation}
\Delta_{ij}=-\frac{1}{2\sqrt{2}y^2}\left(
\begin{array}{ccc}
0 & 2 & 0 \\
2 & 0 & y \\
0 & y & 0 \\
\end{array}
\right).
\end{equation}
The eigenvalues of ${\Delta^i}_j = g^{ik}\Delta_{kj}$ are 0, $\frac{1}{2}$, and $-\frac{1}{2}$.  The eigenvalue of 0 corresponds to the director, because the biaxial splay tensor is constructed to be orthogonal to the director.  The other two eigenvalues correspond to the eigenvectors $u^i=(-y/2,y/\sqrt{2},0)$ and $v^i=(y/2,y/\sqrt{2},0)$, respectively.  Hence, the director field splays outward in the $\mathbf{u}$ direction, and it splays inward in the $\mathbf{v}$ direction.

This is the first time that we have found a director field with pure constant biaxial splay.  At this point, we have identified non-Euclidean geometries with each pure type of nematic director deformation.   Still, we will continue analyzing the last two Thurston geometries, which will provide further examples.

\subsection{$Nil$ space}

$Nil$ geometry is a homogeneous 3D space associated with the Heisenberg matrix group, the multiplicative group of triangular real matrices of the form~\cite{Scott1983,Molnar2003}
\begin{equation}
H=\left(
\begin{array}{ccc}
1 & x & z \\
0 & 1 & y \\
0 & 0 & 1 \\
\end{array}
\right).
\end{equation}
Such matrices can be labeled by the triplet $(x,y,z)$, which provides a simple representation in Euclidean 3D space.  However, $Nil$ geometry is highly non-trivial, since ``translations'' in $Nil$  correspond to matrix multiplication, and are therefore (generically) non-commutative.  A translation by $Nil$ vector $(x,y,z)$ acts on a point $(a,b,c)$, leading to the point $(x+a,y+b,z+c+xb)$, as given by the matrix product
\begin{equation}
\left(
\begin{array}{ccc}
1 & x & z \\
0 & 1 & y \\
0 & 0 & 1 \\
\end{array}
\right).
\left(
\begin{array}{ccc}
1 & a & c \\
0 & 1 & b \\
0 & 0 & 1 \\
\end{array}
\right)
=
\left(
\begin{array}{ccc}
1 & x+a & z+c+x b \\
0 & 1 & y+b \\
0 & 0 & 1 \\
\end{array}
\right).
\end{equation}
$Nil$ space has the covariant metric tensor
\begin{equation}
g_{ij}=
\left(
\begin{array}{ccc}
1 & 0 & 0 \\
0 & 1+x^2 & -x \\
0 & -x & 1 \\
\end{array}
\right),
\end{equation}
and hence the Ricci tensor
\begin{equation}
R_{ij}=
-\frac{1}{2}\left(
\begin{array}{ccc}
1 & 0 & 0 \\
0 & 1-x^2 & x \\
0 & x & -1 \\
\end{array}
\right).
\end{equation}
The scalar curvature is $R=-\frac{1}{2}$ for this negatively curved space, and the anisotropy is shown by $diag({R^i}_j)=(-\frac{1}{2},-\frac{1}{2},\frac{1}{2})$.

\begin{figure}
\begin{center}
\includegraphics[width=0.5\textwidth]{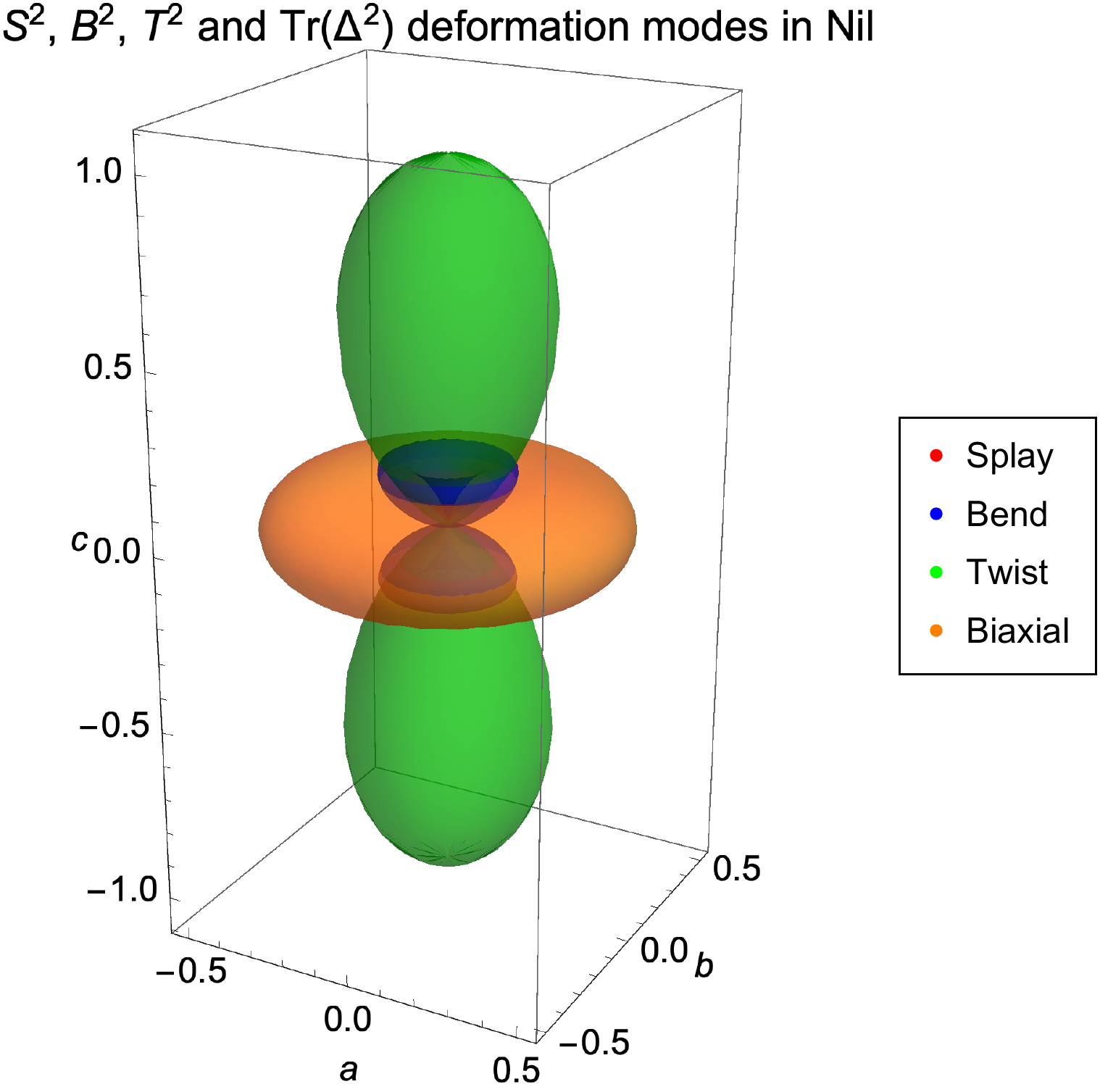}
\end{center}
\caption{Deformation modes for the $Nil$ manifold.  The intensity of each deformation mode is represented by the distance of the corresponding surface from the origin, at that orientation of $\hat{\mathbf{n}}$.}
\label{chemicalnil}
\end{figure}

Once again, the third eigenvalue of the mixed Ricci tensor is distinct from the other two, and hence the corresponding eigenvector must be $c^i=(0,0,1)$.  The other two orthonormal eigenvectors can be chosen as $a^i=(1,0,0)$ and $b^i=(0,1,x)$.  As before, we define the director as $\hat{\mathbf{n}}=\sin \beta \cos \alpha \,\hat{\mathbf{a}} +\sin \beta \sin \alpha\,\hat{\mathbf{b}}+ \cos \beta\,\hat{\mathbf{c}}$ with fixed $\alpha$ and $\beta$.  The deformation modes can then be calculated as
\begin{equation}
S^2 =0,\quad |\mathbf{B}|^2 =\sin^2 \beta\cos^2 \beta, \quad T^2 =\cos^4 \beta,\quad\Tr(\Delta^2)=\frac{1}{2}\sin^4 \beta.
\end{equation}
Figure~\ref{chemicalnil} gives a polar plot of the deformation modes.  All of the director fields in this family satisfy the relation
\begin{equation}
S^2+2|\mathbf{B}|^2+T^2+2\Tr(\Delta^2)=1.
\label{CNil}
\end{equation}
This relation is similar to Eqs.~(\ref{CH2R}) and~(\ref{CSL2R}) for $\mathbb{H}^2\times\mathbb{R}$ and $\widetilde{SL}(2,R)$, except that the relation applies in the full space $Nil$, and the constant is different.

%
%

We now have a situation similar to $\widetilde{SL}(2,R)$, with two directions of negative average sectional curvature, and one with a positive value. Hence, we can test our conjecture that a pure twist configuration may obtained by a director field aligned with the positive direction.  Indeed, we find that behavior:  The director field $\hat{\mathbf{n}}=\hat{\mathbf{c}}$, with $\beta=0$, shows a pure constant twist with $T^2=1$, and zero splay, bend and biaxial splay.

For another interesting case, we consider a director field in the $\{\hat{\mathbf{a}},\hat{\mathbf{b}}\}$ plane, with $\beta=\pi/2$ and arbitrary $\alpha$, so that $n^i=(\cos\alpha,\sin\alpha,x\sin\alpha)$.  This director field has zero splay, zero twist, zero bend, but constant nonzero biaxial splay $\Tr(\Delta^2)=\frac{1}{2}$.  The tensor structure of the biaxial splay is
\begin{equation}
\Delta_{ij}=
\frac{1}{2}
\left(
\begin{array}{ccc}
0 & -x\sin\alpha & \sin\alpha \\
-x\sin\alpha & 2x\cos\alpha & -\cos\alpha \\
\sin\alpha & -\cos\alpha & 0 \\
\end{array}
\right).
\end{equation}
The eigenvalues of ${\Delta^i}_j = g^{ik}\Delta_{kj}$ are 0, $\frac{1}{2}$, and $-\frac{1}{2}$.  The eigenvalue of 0 corresponds to the director, as always.  The other two eigenvalues correspond to the eigenvectors $u^i=(\sin\alpha,-\cos\alpha,1-x\cos\alpha)/\sqrt{2}$ and $v^i=(-\sin\alpha,\cos\alpha,1+x\cos\alpha)/\sqrt{2}$, respectively.  Hence, the director field splays outward in the $\mathbf{u}$ direction, and it splays inward in the $\mathbf{v}$ direction.  This is therefore a new example of a geometry with a pure constant biaxial splay configuration.

\subsection{$Sol$ space}

As discussed by Scott~\cite{Scott1983}, the space $Sol$ has the least symmetry of all eight homogeneous geometries.  Even so, we can follow the same procedure as for the other geometries, and consider simple field configurations in $Sol$.  To parameterize this space, we can use the standard coordinates $(x,y,z)$, with a special role for the $(x,y)$ plane.  A translation has the form~\cite{Scott1983}
\begin{equation}
(x,y,z).(x',y',z')=(x+e^{-z}x',y+e^z y',z+z'),
\end{equation}
with $(0,0,0)$ as the identity.  This corresponds to a standard translation when limited to the $(x,y)$ plane, but is more complicated when leaving that plane.  With this parameterization, the metric becomes $d\ell^2 = e^{2z}dx^2 + e^{-2z}dy^2 + dz^2$. The scalar curvature is $R=-2$ for this negatively curved space, with an anisotropy shown by $diag({R^i}_j)=(0,0,-2)$.

%
\begin{figure}
\begin{center}
\includegraphics[width=0.5\textwidth]{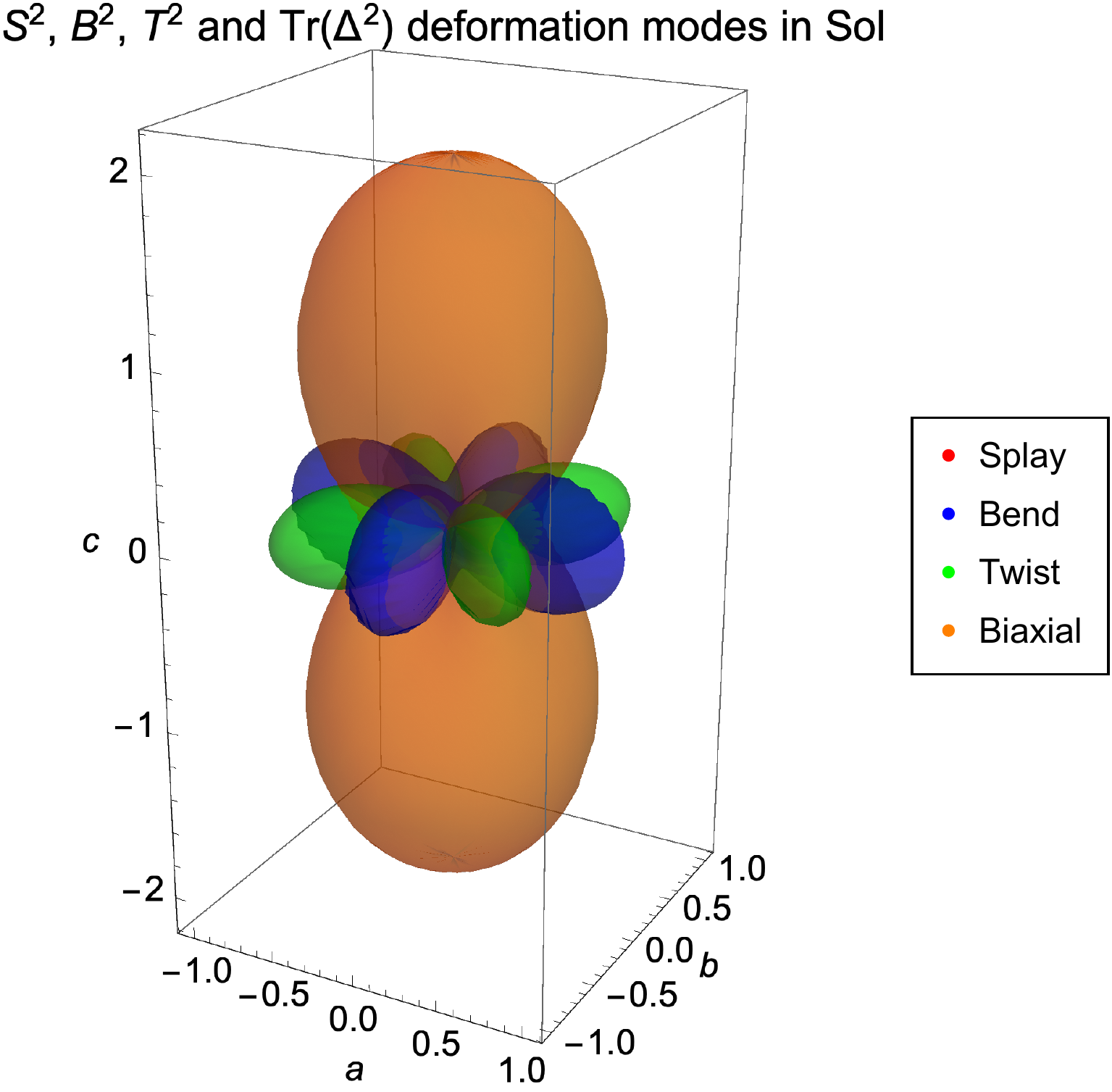}
\end{center}
\caption{Deformation modes for the $Sol$ manifold.  The intensity of each deformation mode is represented by the distance of the corresponding surface from the origin, at that orientation of $\hat{\mathbf{n}}$.}
\label{chemicalsol}
\end{figure}
%

As in several previous cases, the third eigenvalue of the mixed Ricci tensor is distinct from the other two, and the corresponding eigenvector must be $c^i=(0,0,1)$.  The other two orthonormal eigenvectors can be chosen as $a^i=(e^z,0,0)$ and $b^i=(0,e^{-z},0)$.  As usual, we define the director as $\hat{\mathbf{n}}=\sin \beta \cos \alpha \,\hat{\mathbf{a}} +\sin \beta \sin \alpha\,\hat{\mathbf{b}}+ \cos \beta\,\hat{\mathbf{c}}$.  The deformation modes then become
\begin{eqnarray}
S^2 =0,\quad |\mathbf{B}|^2 =\frac{1}{4}\sin^2 \beta\left[3+\cos2\beta+2\cos4\alpha\sin^2 \beta\right],\\
T^2 =2\sin^2 2\alpha\sin^4 \beta,\nonumber\\
\Tr(\Delta^2)=\frac{1}{32} \left[35-3\cos4\alpha+4\cos2\beta(7+\cos4\alpha)+2\sin^2 2\alpha\cos4\beta\right].\nonumber
\end{eqnarray}
These four modes are shown in the polar plot of Fig.~\ref{chemicalsol}.  We cannot find a general relation among the modes that is valid in the entire space, and hence we concentrate on the mixed Ricci eigenplane spanned by $\hat{\mathbf{a}}$ and $\hat{\mathbf{b}}$, where $\beta=\pi/2$.  In that plane, we have
\begin{equation}
S^2+2|\mathbf{B}|^2+T^2+2\Tr(\Delta^2)=2.
\label{CSol}
\end{equation}
This relation is analogous to Eqs.~(\ref{CH2R}), (\ref{CSL2R}), and~(\ref{CNil}) in the spaces $\mathbb{H}^2\times\mathbb{R}$, $\widetilde{SL}(2,R)$, and $Nil$, respectively.

We can now analyze properties of some simple configurations in $Sol$.  First, consider the director field $\hat{\mathbf{n}}=\hat{\mathbf{c}}$, i.e.\ $\beta=0$, which is tangent to lines of varying $z$ with constant $x$ and $y$.  This director field has pure constant biaxial splay $\Tr(\Delta^2)=2$, with zero splay, twist, and bend.  The tensor structure of the biaxial splay is
\begin{equation}
\Delta_{ij}=
\left(
\begin{array}{ccc}
e^{2z} & 0 & 0 \\
0 & -e^{-2z} & 0 \\
0 & 0 & 0 \\
\end{array}
\right).
\end{equation}
The eigenvalues of ${\Delta^i}_j = g^{ik}\Delta_{kj}$ are $0$, $1$, and $-1$.  These eigenvalues correspond to the director (along $z$) and to the $x$ and $y$ directions, respectively.  Hence, the director field splays outward in the $x$ direction, and it splays inward in the $y$ direction.

Another interesting class of director fields is in the $\{\hat{\mathbf{a}},\hat{\mathbf{b}}\}$ plane, with $\beta=\pi/2$ and arbitrary $\alpha$, so that $n^i=(e^{-z} \cos\alpha,e^z\sin\alpha,0)$.  This class of director fields has zero splay, constant twist $T^2 =\sin^2 2\alpha$, constant bend $|\mathbf{B}|^2 = \cos^2 2\alpha$, and constant biaxial splay $\Tr(\Delta^2)=\frac{1}{2}\sin^2 2\alpha$.  If $\alpha=0$ or $\frac{\pi}{2}$, the director field is tangent to lines of varying $x$ or $y$, respectively, and it has pure constant bend, with zero splay, twist, and biaxial splay.

\section{Conclusions}

In this article, we have explored the eight Thurston 3D geometries, which are all the possible homogeneous 3D spaces, and we have investigated whether they allow director field configurations with pure constant deformation modes.  This study can be viewed as a generalization of the work of Sethna \emph{et al.}~\cite{Sethna1983} for the double twist case on $\mathbb{S}^3$.  We were indeed able to find several such configurations, which we summarize here:

\begin{itemize}

\item\emph{Pure constant splay}

We found a 2D example in $\mathbb{H}^2$, and a 3D example in $\mathbb{H}^3$.

\item\emph{Pure constant bend}

We found a 2D example in $\mathbb{H}^2$, and 3D examples in $\mathbb{H}^3$, ${\mathbb{H}^2 \times \mathbb{R}}$, and $Sol$.

\item\emph{Pure constant twist}

In addition to the previously known example in  $\mathbb{S}^3$, we found new examples in $\widetilde{SL}(2,R)$ and $Nil$. Notice that although these two geometries have negative scalar curvature, they each have a direction such that the average sectional curvature, for planes sharing this direction, is positive.  In each case, the double twist solution is aligned with that direction.

\item\emph{Pure constant biaxial splay}

We found examples in $\widetilde{SL}(2,R)$, $Nil$ and $Sol$.

\end{itemize}

Based on these examples, we can see that compatibility between director deformations and spatial geometry is much more complex in 3D than in 2D.  In 2D, Niv and Efrati~\cite{Niv2018} derived the simple relation $S^2 + B^2 = -K_G$ between constant splay $S$, constant bend $B$, and the Gaussian curvature $K_G$ of the surface.  This relation shows that a nonzero splay or bend can exist everywhere in a 2D surface with the appropriate negative curvature, but must be frustrated in a surface with zero or positive curvature.  By contrast, in 3D, we find that the compatibility of a director field depends on the type of director deformation and on the type of curved geometry.  In most cases we could nevertheless find that, for director fields which have constant orientation in the tangent space derived from a global coordinate system,  the expression $S^2+2|\mathbf{B}|^2+T^2+2\Tr(\Delta^2)$ is constant in sub-manifolds which are eigenspaces of the mixed Ricci tensor. Interestingly, the latter expression is proportional to the one-constant free energy expression given in Eq.~(\ref{equa8}). The corresponding sub-manifolds where this expression is constant are therefore degenerate with respect to this form of the free energy (which is only one among the different forms of free energy discussed in Sec.~2.3).

One might hope that the examples found here will guide future work on the general compatibility conditions, indicating what mathematical constraints must be satisfied for a deformation to fill up a space.

In any case, these examples certainly show that the formation of an ideal liquid-crystal deformation is a subtle geometric issue.  In ordinary Euclidean space $\mathbb{R}^3$, such structures are \emph{usually} frustrated.  For that reason, liquid crystals \emph{often} form extra deformation modes beyond the favored mode (as in the heliconical twist-bend nematic phase), or form lattices of defects (as in blue phases).  As a result, this geometric frustration leads to rich phase behavior of liquid crystals.

\section*{Acknowledgments}

We would like to thank E.~Efrati and P.~Pansu for helpful discussions.  JVS was supported by National Science Foundation Grant DMR-1409658.  Part of this work was performed at the Aspen Center for Physics, which is supported by National Science Foundation Grant PHY-1607761.

\section*{References}

\bibliographystyle{iopart-num}
\bibliography{NonEuclideanLCv7}

\end{document}